\documentclass[traditabstract]{aa} 
\usepackage{longtable}
\usepackage{latexsym}
\usepackage{amssymb}
\usepackage{lscape}
\usepackage[]{natbib}

\usepackage{color}

\usepackage{graphicx}
\usepackage{txfonts}
\usepackage{float}
\def\lsim{\mathrel{\rlap{\lower 3pt \hbox{$\sim$}} \raise 2.0pt \hbox{$<$}}}
\def\gsim{\mathrel{\rlap{\lower 3pt \hbox{$\sim$}} \raise 2.0pt \hbox{$>$}}}

   \title{Robust automatic photometry of local galaxies from SDSS}
   \subtitle{Dissecting the color magnitude relation with color profiles}
   \author{Guido Consolandi \inst{1}
   \and Giuseppe Gavazzi \inst{1}
   \and Michele Fumagalli \inst{2}
   \and Massimo Dotti \inst{1}
   \and Matteo Fossati  \inst{3\and4}
   }
   \authorrunning{G. Consolandi et al.}
   \institute{
    Dipartimento di Fisica G. Occhialini, Universit\`a di Milano-Bicocca, Piazza della Scienza 3, I-20126 Milano, Italy\\
    \email{guido.consolandi@mib.infn.it, giuseppe.gavazzi@mib.infn.it, massimo.dotti@mib.infn.it} 
    \and
   Institute for Computational Cosmology and Centre for Extragalactic Astronomy, Department of Physics, Durham University, South Road, Durham DH1 3LE, UK \\
    \email{michele.fumagalli@durham.ac.uk} 
    \and 
    Universit{\"a}ts-Sternwarte M{\"u}nchen, Scheinerstrasse 1, D-81679 M{\"unchen}, Germany\\	     
    \email {mfossati@mpe.mpg.de}
    \and
    Max-Planck-Institut f{\"u}r Extraterrestrische Physik, Giessenbachstrasse, D-85748 Garching, Germany
}
\begin{document}
\date{}

  \abstract{ We present an automatic procedure to perform reliable
    photometry of galaxies on SDSS images. We selected a sample of
    5853 galaxies in the Coma and Virgo superclusters. For each galaxy, we
    derive Petrosian $g$ and $i$ magnitudes, surface brightness
    and color profiles.  Unlike the SDSS pipeline, our procedure is
    not affected by the well known shredding problem and efficiently extracts
    Petrosian magnitudes for all galaxies. Hence we derived  magnitudes even from the population of galaxies missed 
    by the SDSS which represents $\sim 25\%$ of all Local supercluster galaxies and $\sim 95\%$ of galaxies with $g < 11$ mag. After correcting the
    $g$ and $i$ magnitudes for Galactic and internal extinction, the blue
    and red sequences in the color magnitude diagram are well separated, with
    similar slopes. In addition, we study (i) the color-magnitude diagrams
    in different galaxy regions, the inner (r $\le 1$ kpc), intermediate
    (0.2R$_{Pet} \le$r$\le 0.3$R$_{Pet}$) and outer, disk-dominated
    (r$\ge0.35$R$_{Pet}$)) zone, and (ii), we compute template color profiles,
    discussing the dependences of the templates on the galaxy masses and on
    their morphological type. The two analyses consistently lead to a picture
    where elliptical galaxies show no color gradients,
    irrespective of their masses. Spirals, instead, display a steeper gradient
    in their color profiles with increasing mass, which is consistent with the
    growing relevance of a bulge and/or a bar component above $10^{10}\rm M_\odot$.}
   

\maketitle


\section{Introduction}

The advent of the Sloan Digital Sky Survey (SDSS, York et al. 2000) represents
a revolution in many fields of observational astronomy.
The final SDSS
catalog consists of about 500 million photometric objects in five bands with
more than 1 million spectra.  This huge amount of publicly available data has
given the community the unprecedented opportunity to study galaxy
properties from a statistical point of view in many different regimes of local
density and cosmic epochs.

Early multi-wavelength studies by  \citet{faber73}, \citet{visva77}, 
\citet{aar81} suggested the existence for E 
and S0s of a color-magnitude relation with a shallow slope and a small intrinsic
scatter. Similarly, other authors stated that spiral galaxies follow a color 
magnitude relation with a much greater intrinsic scatter \citep{ches64,visva77b,grier80}. 
However these pioneering studies were based on limited samples of only  tens or at most hundreds of
objects  which prevented conclusive statistical analyses.
It was only with the advent of the SDSS that \citet{strat01},
using an unprecedented sample of 147920 galaxies, have definitely demonstrated
the bimodal distribution of galaxy colors and luminosity, with a 
clear separation between star-forming blue late-type galaxies and passive, 
red and dead galaxies. 
Since 2001, this bimodality in galaxy distribution has been
observed in many different regimes of density \citep{bal04,kauff04,pg10} at
different redshifts \citep{bel04, hzcmag} suggesting an evolutionary path of galaxies from the so-called blue
cloud of star forming galaxies to the red sequence of dead galaxies.  However, the 
mechanisms that make galaxies depart from the blue cloud to reach the red
sequence (star formation-quenching) are still under scrutiny.

\begin{figure*}
\begin{centering}
\includegraphics[width=15cm]{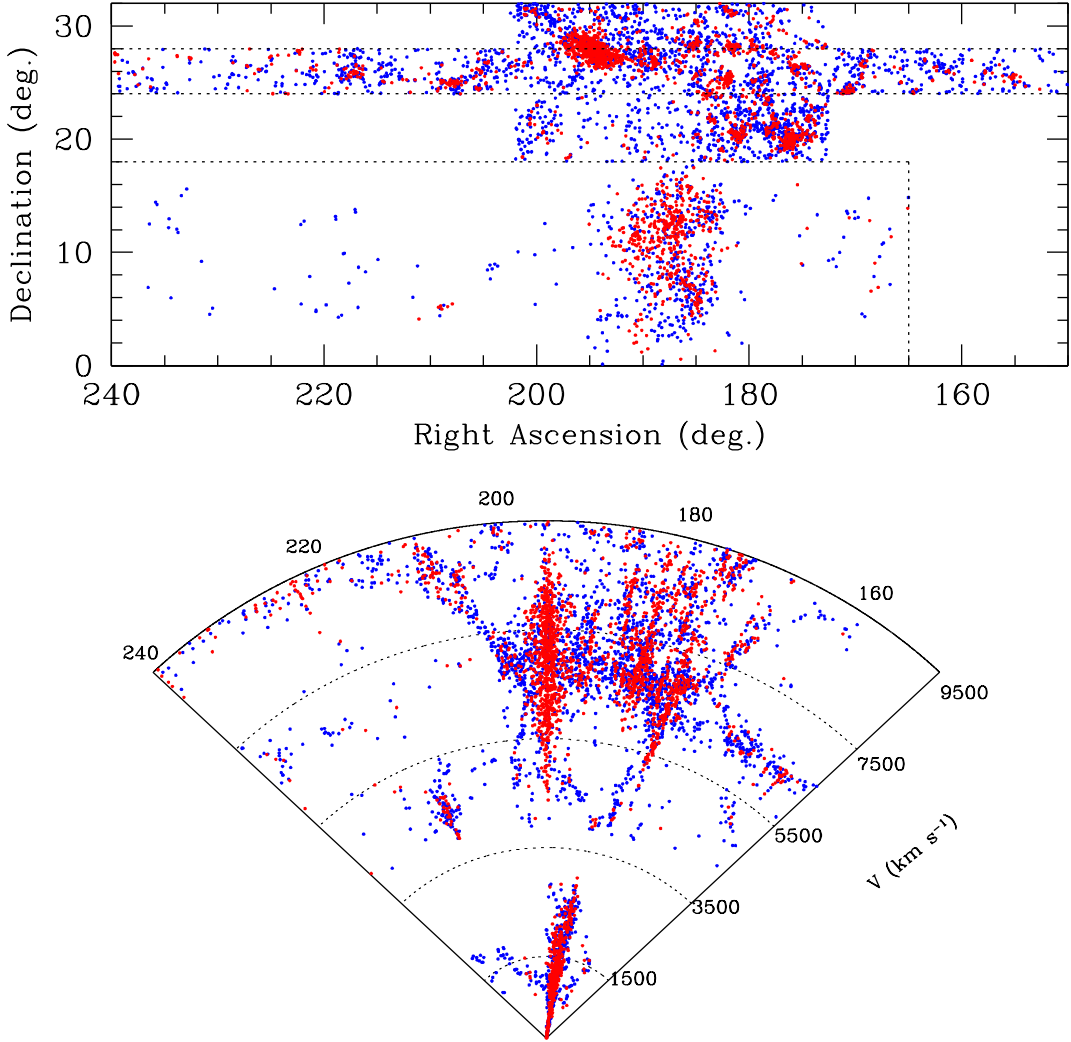}
\caption{Sky projection (top) and Wedge diagram (bottom) of galaxies belonging to the sample studied in this work:
the Coma supercluster ($\delta>18^o$; $cz>4000$ km~s$^{-1}$) and the Local supercluster ($\delta<18^o$; $cz<3000$ km~s$^{-1}$).
Blue dots represent late yype Galaxies (LTGs) while red dots stand for early type galaxies (ETGs). 
The dotted rectangular regions indicate the areas surveyed by ALFALFA.}
\label{sample}
\end{centering}
\end{figure*}

However 
the SDSS pipeline was optimized for the extraction of
photometric parameters of galaxies at the median redshift of the survey ($z \sim 0.1$).
Large nearby galaxies, often exceeding the apparent diameter of few arcmins
were obviously penalized by this choice.  One longstanding  problem that 
affects the identification of extended sources \citep{blanton05} 
is the so-called shredding of large, bright  (in apparent magnitude) galaxies by the automatic pipelines,
leading to wrong detections and incorrect magnitude calculations.
Moreover, restrictions due to fiber collisions dictate that no two fibers can be placed closer 
than $55$ arcsec during the same observation \citep{blanton03} which
affects the spectroscopy of crowded environments and introduces incompleteness at $z<0.03$ 
where many local large scale structures exist, namely the Local and the Coma superclusters (see Fig. \ref{sample}).

Nevertheless, galaxies of the local universe are, at fixed luminosity, the brightest as well as the best spatially resolved objects 
of the sky. Therefore, within the SDSS, local galaxies can be analyzed down to a mass limit that cannot be reached at higher redshift and,
in addition, their structures and  morphology  are easier to constrain.
In fact the median resolution of the SDSS, given by the  median point spread function (PSF  $\sim1.4$ arcsec, in the $r$-band), corresponds to a physical 
scale of $\sim 600$ pc at the distance of the Coma cluster ($\sim 95.38$ Mpc) and $\sim 80$ pc at the distance of 
Virgo \citep[$\sim 17$ Mpc,][]{pg99,mei07}.  This resolution is  
sufficient to resolve even small structures such as nuclear disks in galaxies.\\
Henceforth local galaxies are a natural  test-bed for studies on the evolution of structures at higher redshift. 
The Local and Coma superclusters contain thousands of galaxies and are a perfect laboratory for  studying the leading processes that
transform galaxies from star forming objects into red and dead structures.
Moreover   the nearby Universe contain a wide range of environments that have been broadly studied by many groups 
at different wavelengths \citep{alfa40,guvics, pg10, pg12}.  These regions are  therefore perfect to constrain the properties 
of galactic structures in different environments at different wavelengths, with the SDSS magnitudes playing a fundamental role in 
the panchromatic description of these structures. 
 Both areas of the sky have been fully covered by the SDSS in its five optical filters ($u$, $g$, $r$, $i$, $z$) and the data have 
been first published in the data release 4 (DR4) for \citep[Virgo,][]{dr4} and in the DR7 \citep[Coma,][]{dr7}.
Unfortunately, the data of the  local and Coma clusters as given by SDSS are affected by the aforementioned 
difficulties of the SDSS pipeline and, despite the good quality of the data, many galaxies are  not included in the
catalogs or, if present,  have in some cases unreliable photometry. 
 Despite the fact that the worst photometric discrepancies affect only $\sim10\%$ of the whole sample (see section \ref{mags}),
these happen to coincide with the brightest galaxies in the Virgo and Coma clusters, thus affecting the  high mass-end determination of the luminosity function that is 
already hampered by the lack of sampled volume.
 
In this context, our work aims to automatically generate high-quality photometry from SDSS imaging in two filters ($g$ and $i$)
for a sample of $\sim 6000$ galaxies within the Local and Coma superclusters described in section 2. Our IDL-based (Interactive Data Language)  procedure is not limited to the 
magnitude extraction but performs aperture photometry in each considered band as well.
In section 3 we explain the method developed while in section 4 we compare our magnitudes with the SDSS database, the Extended 
Virgo Cluster Catalog \citep{Kim+14} and measurements from \citet{pg12} and \citet{pg13b}. We test quantitatively 
the efficiency of the SDSS pipeline in these local clusters, while demonstrating
the exquisite quality of our  guided extraction. Section 5, 6 and 7 focus on the color-mass distribution 
of galaxies within our sample and on the properties of the different galactic components traced by color profiles as a function 
of mass and morphology. Our findings are discussed and summarized in section 8.

\section{The sample}
\label{secsample}
This work is based on a sample of 6136 nearby galaxies in the spring sky selected from the SDSS  as described in this section. The final sample is further split in two subsamples: 
i) the Local Supercluster ($11^h<RA<16^h$; $0^o<Dec<18^o$; $cz<3000 ~\rm km ~sec^{-1}$) 
containing 1112 galaxies and includes the Virgo cluster; 
ii) the Coma Supercluster ($10^h<RA<16^h$; $18^o<Dec<32^o$; $4000<cz<9500 ~\rm km ~sec^{-1}$) containing 5024 galaxies
and includes the Coma cluster. The two subsamples are displayed in Figure \ref{sample}.
\begin{figure}
\begin{centering}
\includegraphics[width=9.5cm]{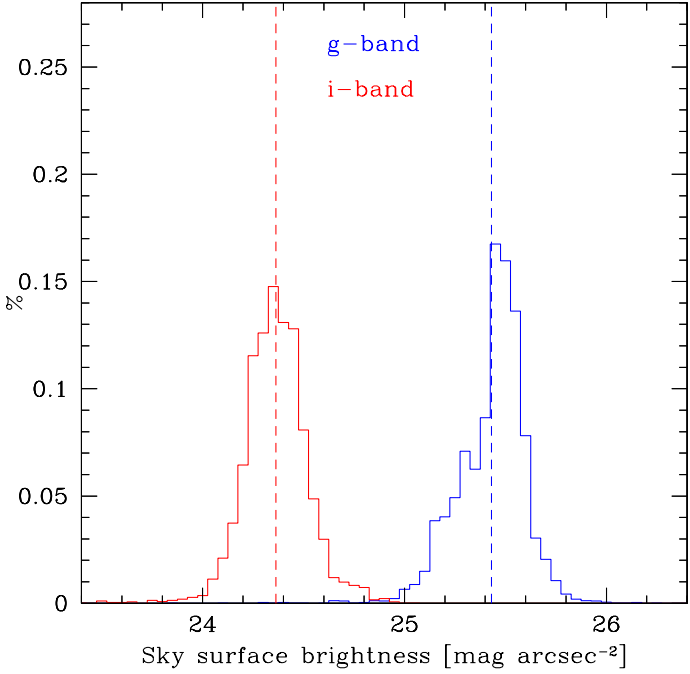}
\caption{Sky surface brightness distributions in $g-$ (blue line) and $i-$ band (red line) for the sample analyzed in this work. The vertical dashed lines indicate
the mean surface brightness in $g-$ and $i-$ band respectively $<\Sigma_i>=24.36\pm0.16$ and $<\Sigma_g>=25.43\pm0.17$ mag arcsec$^{-2}$. }
\label{SB}
\end{centering}
\end{figure}

Since galaxies at the distance of Virgo have apparent size often exceeding 5 arcmin, they are strongly affected by 
the shredding problem (Blanton et al. 2005) therefore our catalog cannot solely rely on the SDSS spectroscopic database. Thus, the Local supercluster sample is selected following the prescriptions of \citet{pg12}:
in the area occupied by the Virgo cluster,the selection is based on the VCC catalog (limited however to $cz<3000 \rm
~km ~sec^{-1}$) down to its magnitude completeness limit of 18 mag \citep{vcc}. The object selection is furthermore limited to objects with surface brightness above the
$1\sigma$ of the mean sky surface brightness in $i-$ band 
of the SDSS data (see Fig. \ref{SB}). Outside the Virgo cluster the SDSS selection is complemented with objects taken from NED and ALFALFA \citep{alfa40}. 

At the distance of the Coma supercluster the shredding problem is less severe and therefore we followed the
selection of \citet{pg10, pg13b}. Briefly, galaxies are selected from the SDSS spectroscopic database DR7 \citep{dr7} with $r<17.77$ mag. To fill the residual incompleteness of the SDSS catalog for extended galaxies
and due to fiber conflict,  133 galaxies from the CGCG catalog \citep{cgcg} with known redshifts from NED and 28
from ALFALFA are added, reaching a total of 5024 galaxies.

For these galaxies, we downloaded the SDSS images (using the on-line Mosaic service, see next Section) 
in the $g-$ and $i-$band. In this work, we test the procedure exclusively on these two bands for mainly two reason:
$i$) The u and z filters have lower signal-to-noise ratio (SN), $ii$) if compared to the $i-$band, the $r$ filter have a central wavelength closer to the $g$ filter central wavelength.
Hence the $g$-$i$ color is more sensitive to stellar population gradients and dust absorption.
However only for 5753 (94\%) targets the download process worked in both the $i-$ and the $g-$band
\footnote{the procedure is able to process all the SDSS bands images at the same time.
Nevertheless, in this paper we present the analysis of only the $g-$ and $i-$band images. Obviously adding more bands  affects 
the SN of the white image and therefore the detection performed by Source Extractor}.
Of these, 221 galaxies were discarded a posteriori because they lie too close to bright stars or they have too low 
surface brightness (see Fig. \ref{SB}).
The remaining analyzed sample is constituted of 969 (Local SC) + 4563 (Coma SC) objects,  for a total of
5532 galaxies that can be considered representative of the nearby universe. Their morphological classification and diameters are 
taken from the public database GOLDMine \citep{pg03,pg14b}. 

Summarizing, the sample analyzed in this work coincides with the one presented by \citet[][with the selection criteria given in Gavazzi et al. 2010, Coma supercluster]{pg13b} 
and Gavazzi et al 2012 (Local supercluster) except 
for 108 VCC galaxies that were excluded because their surface brightness  (evaluated inside the radius at the $25^{th}$ B-band isophote reported in GOLDMine) is lower than
the  mean sky surface brightness in $i-$ band (Fig. \ref{SB}). For the remaining  
5532  galaxies the procedure developed in this work is aimed at obtaining more
accurate photometry than reported in Gavazzi et al. (2012, 2013b).
\section{The method}
\label{proc}
Our analysis code has been designed using IDL with the IDL Astronomy User's Library 
(Landsman 1993) and takes advantage of Source Extractor (Bertin \& Arnouts 1996). It processes
\emph{FITS} SDSS images in multiple bands that were downloaded using the
IRSA and NVO image Mosaic service \citep{montage,k11}.   The Mosaic interface returns
science-grade SDSS mosaics that preserve fluxes and astrometry and rectify
backgrounds to a common level.  Furthermore, the photon counts are normalized to
a common exposure time of 1 s and the zero point is  set to 28.3 mag in every filter.  
Images are centered on the target galaxies and span approximately
three times their major diameter at the $25^{th}$ magnitude ${\rm arcsec^{-2}}$ V-band isophote as reported in 
GOLDMine \citep{pg03,pg14b}. This ensures that a sufficient number of sky pixels exist around the targets 
for robust measurement of the sky background.\\

\subsection{Image preparation, target detection and masking}
\label{phot_reduction}

The first step of the procedure is  a preliminary estimate of the sky in each filter 
in a rectangular peripheral corona of width equal to one quarter of the full image size. 
This  background mode value is subtracted from the images.
The $g$ and $i$ sky-subtracted images are then averaged to create a higher signal-to-noise white frame 
that is analyzed using Source Extractor which detects and determines the photometric and geometric parameters 
of the objects in the field. We facilitated the Source Extractor detection enabling the filtering option of the routine.
Each image is filtered with a Gaussian filter in order to smooth the image and avoid the detection of bright substructure inside the target galaxy as well as 
help the detection of the low surface brightness Local supercluster irregular galaxies.
In the Source Extractor setup that we adopted (for the technical details about the implementation we refer the  reader to Appendices~\ref{setup}~and~\ref{bigdeal})
 we apply a 13x13 pixels smoothing filter. This value is obtained after that we tested that a finer filter does not cure the shredding problems while a larger filter
drastically affects the deblending of overlapping objects.\\

Source Extractor succeeds in identifying the target galaxy as the central object in $96\%$ of the cases.
It also identifies all other objects in the field (overlapping or not with the target),
discriminating between stars and galaxies through a continuum $CLASS\_STAR$
parameter that runs form 0 (galaxies) to 1 (stars).   In this work, a lower limit of 0.8 has
been adopted for an object to be considered a star. 
A  mask of all the sources is then built and, in order to prevent from masking substructures inside  a galaxy that are erroneously detected as overlapping 
sources, our procedure exploits the geometric parameters of the central galaxy as extracted by Source Extractor and define the area that it occupies.
Henceforth we remove masks that are completely embedded within the central galaxy Kron radius \citep{kron}and with a 
$CLASS\_STAR$ parameter lower than 0.8 (non stellar). This avoids that structures such as spiral arms or bright HII regions are masked. 
Despite this geometric criterion, the masking of other sources with partial overlap with the main galaxy is preserved.
Overall, this method fails to work in less than 4\% of all cases, which, after  inspection of the individual cases, 
are found to belong to three different classes:

a) 45 large (A$>$5 arcmin) galaxies ($\sim 1\%$ of the sample) are still 
affected by a serious shredding problem.  To overcome this problem 
we define an elliptical region for the target galaxy whose parameters (major, minor axes and PA ) are
taken from the UGC or the VCC. Within such elliptical shape Source Extractor finds and masks stars;

b) 116 ($\sim {2 \%}$) targets suffer from insufficient masking from the halo, or spikes of bright stars in the field, or 
from the light of companion galaxies. These objects are masked manually;

c) In $\sim 1\%$ if cases, the target galaxies do not lie at the center of their respective
images downloaded with Mosaic. This happens when the target coordinates were
inaccurate or when Mosaic returns images displaced from the nominal
coordinates. In these cases our code masks erroneously the galaxy as it does
not recognize it as the primary target. These cases are repaired by manually
cutting the frames around the target galaxy. 

Altogether cases a+b+c,  the only that requires human intervention, affect 220 objects out of 5532, a mere 4\% of all  cases. 
The remaining 96\% are treated automatically. In this sense our procedure cannot be defined fully automatic, but quasi-automatic 
or guided. 
Once the mask is obtained the sky is re-computed (and re-subtracted) in order to remove the possible contamination by the sources within the corona used for preliminary background subtraction. 
\subsection{Petrosian radius and photometric extraction}
\label{phot}

\begin{figure}
\begin{centering}
\includegraphics[width=9.5cm]{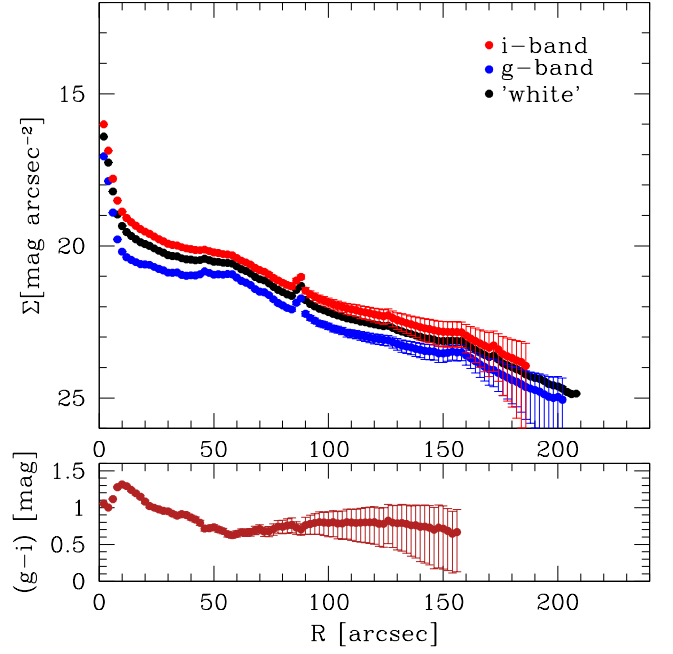}
\caption{Surface brightness profiles of the Virgo galaxy VCC-508. The $g$, $i$ 
and white profiles are respectively show in blue, red and black 
in the top panel. All are traced down to $\Sigma_{{\rm sky}}$. In the bottom panel, the color profile is plotted.}
\label{profil}
\end{centering}
\end{figure}

For each target galaxy  identified in the white image, Source Extractor provides a set of  parameters
(i.e., the center XWIN, YWIN, the position angle PA and the axis ratio B/A).
Using these parameters or, in case of Source Extractor failure, those taken at the 25th magnitude isophote in B-band from GOLDmine (Gavazzi et al 2003, 2014)
our procedure creates a set of concentric ellipses centered on the object and  
oriented at fixed PA, with a constant axis ratio B/A,  evaluated at the 1.5 $\sigma$ of the sky isophote by Source Extractor.
In this work we choose to keep constant axis ratio and P.A. in order to keep a constant radial step between the elliptical annuli avoiding jumps due to
the twist and overlap of the isophotes  in galaxies hosting non-axisymmetric structures such as spiral arms or bars \citep{minc13}. 
\footnote{To speed up the procedure we set the step between adjacent
  ellipses (along the semi-major axis)  as a function of the image dimensions: 1 pixel for images 
  smaller than 300 pixels; 2 pixels for images greater than 300 pixels, up to 1000 pixels; 
  4 pixels for images greater than 1000 pixels.} The concentric ellipses define the
elliptical annuli over which the surface brightness
profile is evaluated as a function of  the distance along the major axis, down to the 
$\Sigma_{sky}$ surface brightness limit.  Within the ellipses, the procedure evaluates the Petrosian
radius $r_{p}$.  This is defined as the radius at which the Petrosian ratio $R_{p}$, defined
as
\begin{equation}
R_{p}(r_{p})=\left(\frac{\int_{0.8r}^{1.25r} dr' 2\pi r' I(r')}{[\pi (1.25^2-0.8^2)r^2]}\right) \left( \frac{\int_{0}^{r}dr' 2\pi r' I(r')}{\pi r^2}\right)^{-1}\:,
\end{equation}
\citep{bla01a, yasu01}
reaches 0.2, in line 
with the value adopted by the SDSS pipeline. The Petrosian flux 
($F_{P}$) is defined as the flux within  $N_p$  Petrosian radii.  We set $N$=2, 
once again consistently  with the SDSS algorithm. 
 For consistency, Petrosian magnitudes are  computed in both the $g$ and $i$ images within the   
the same aperture, determined in the white frame.\\ Similarly, the
set of ellipses generated on the white image is used to evaluate both the $g$
and $i$ surface brightness profiles.  Both are  computed down to the $ \Sigma_{sky}$
surface brightness limit, see the  example shown in Fig. \ref{profil}.\\ 
 Finally, the $(g-i)$ color image is obtained performing the operation 
$Im_{(g-i)}=-2.5\log{\frac{Im_{g}}{Im_{i}}}$, where Im$_{g}$ and Im$_{i}$
are respectively the $g$ and $i$ sky-subtracted image and Im$_{(g-i)}$ is the color
image. The color profile, truncated where the rms $\sim 1\sigma_{sky;(g-i)}$, is obtained subtracting the $i$- from the $g$-band surface brightness profile.

\subsection{Errors}

In order to evaluate errors on the surface brightness profiles,  we take into
account the Poissonian statistical noise ($\sqrt{S}$, where S are the counts)
and the noise contributed by the statistics of  the background,
which has units of flux per area (i.e., surface brightness). This corresponds
to the total observed individual pixel background $\sigma$ integrated over the area A.
Nevertheless, the MONTAGE software resamples the images of multiple SDSS fields building a mosaicked 
image of the target object with a generalized drizzle algorithm. The draw-backs are a  Moir\`e pattern \citep{cotini13}, a slight image degradation \citep{blanton11}and 
a mild correlation in the noise of pixels which can lead to 
an underestimate of the statistic of background. The Moir\'e pattern is in general non relevant \citep{cotini13} and
the degradation of the image is indeed very small \citep{blanton11}. Hence, following standard procedures (e.g., \citet{gawiser} and \citet{fuma14}), we estimated the correlated noise 
that possibly affects our sky 
noise determination, especially within wide apertures. We compute an empirical noise model for the images, by measuring the 
flux standard deviation in aperture of n$_{pix}$ pixels within sky regions empty of sources, according to the segmentation map.
We fit the size-dependent standard deviation with the function
\begin{equation}
\label{sly}
\sigma(n_{pix})=\sigma_1\alpha n_{pix}^{\beta},
\end{equation}
where $\sigma_1$ is the individual pixel standard deviation and $\alpha$ and $\beta$ are the best-fit parameters that typically
have a value of $\alpha\sim 1$ and $\beta\sim0.56$ within our sample (where $\beta=0.5$ means uncorrelated noise), implying a small but non-zero correlation in the pixel noise. 
The sky errors are therefore evaluated following the fitted function.
Finally we consider an additional important source of error that comes from residual gradient of the flat
field, which is estimated to be $10\%$ of the {\rm rms} in the individual
pixel by \citet{pg93}.  This represents the dominant source of error in the low
surface brightness regions and must therefore be taken into account.  {\rm The total}
noise (N) is therefore estimated to be the quadratic sum of the  the empirical noise model, sky  (corrected for the noise correlation)
and flat fielding errors:
\begin{equation}
\label{noise}
N=\sqrt{\frac{C}{G}+(\sigma^2)(\Delta A)^{1.12}+0.1\sigma^2A},
\end{equation}
where C are the counts, G the gain of the CCD, 
$\Delta A$ is the annuli area and A is the area of the aperture.\\
The magnitudes into the aperture are 
\begin{equation}
m \pm \delta(m) = Zp - 2.5\log(S \pm N).
\end{equation}
Errors on magnitudes in the profile can therefore be written as:
\begin{equation}
\label{errr}
\delta m = -2.5\log(1\pm \frac{N}{S}),
\end{equation}
where S is the observed flux and N is the noise. 
In  the ($g-i$) color profiles,  the noise ($N(g-i)$) 
is  computed as the quadratic sum of the $N_g$ and $N_i$ which are the noise computed with equation \ref{noise} respectively in $g-$ and $i-$band.
\section{Results}

\begin{figure*}
\begin{centering}
\includegraphics[width= 7.5cm]{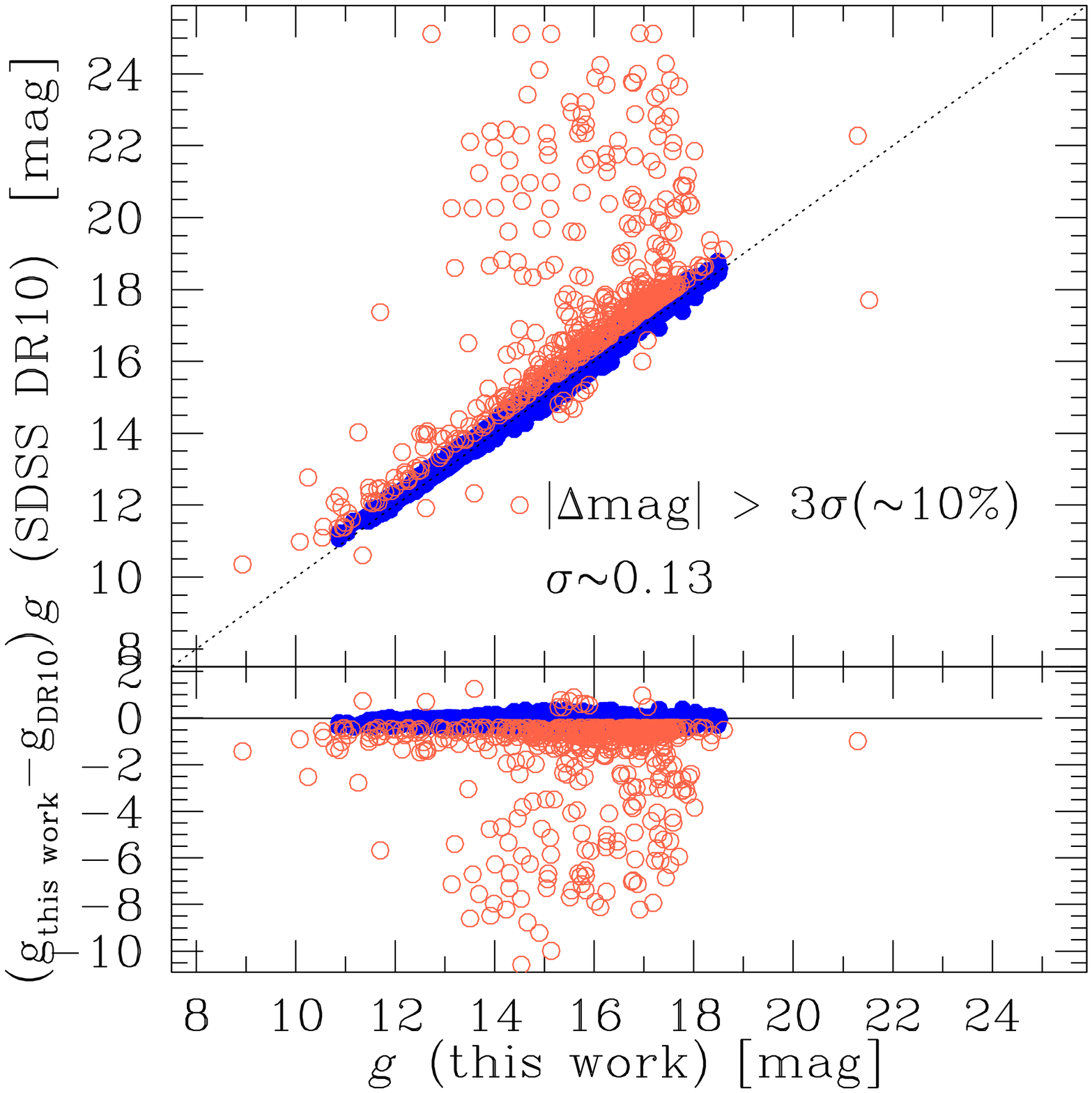} \includegraphics[width= 7.5cm]{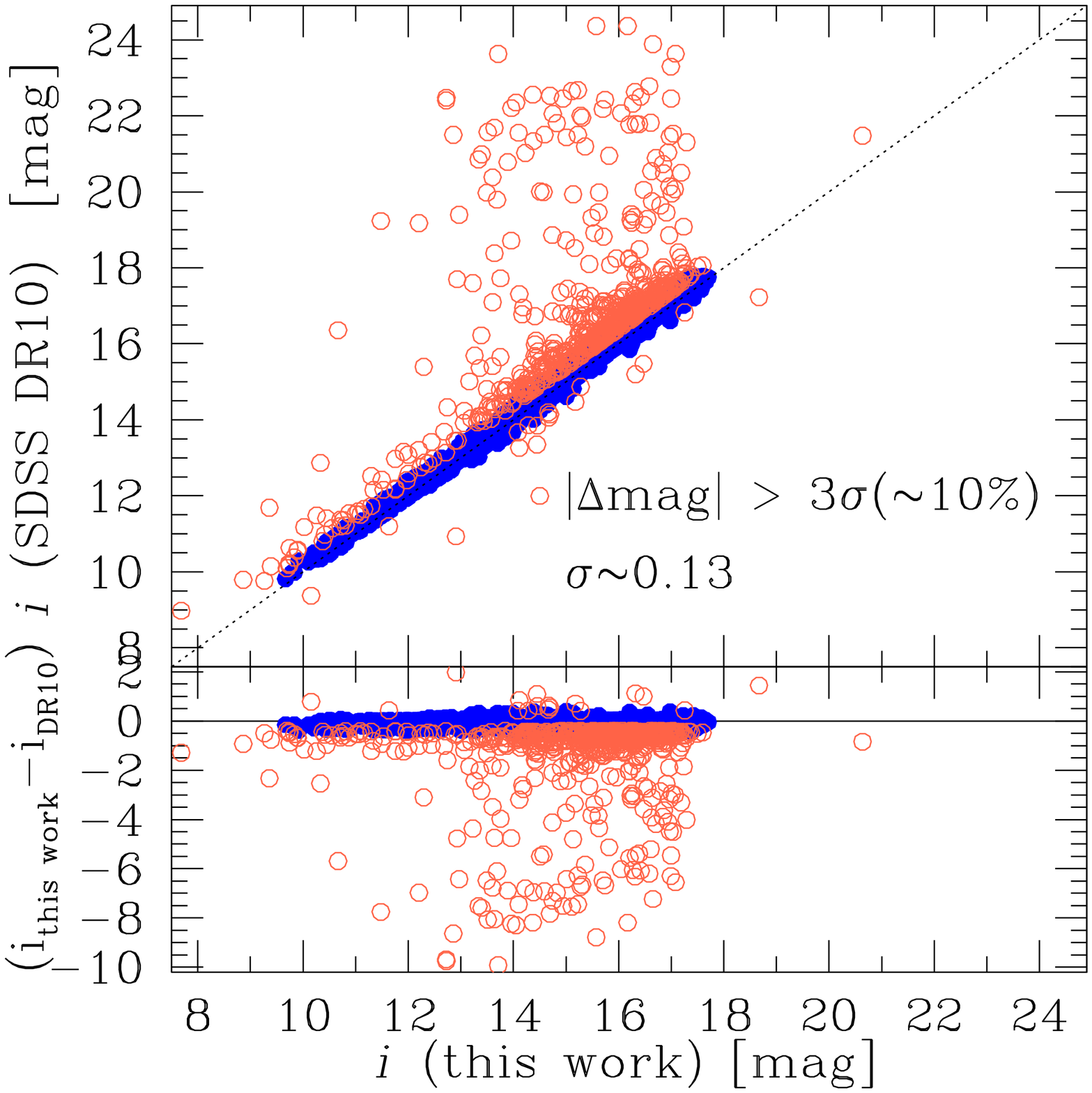}
\includegraphics[width= 7.5cm]{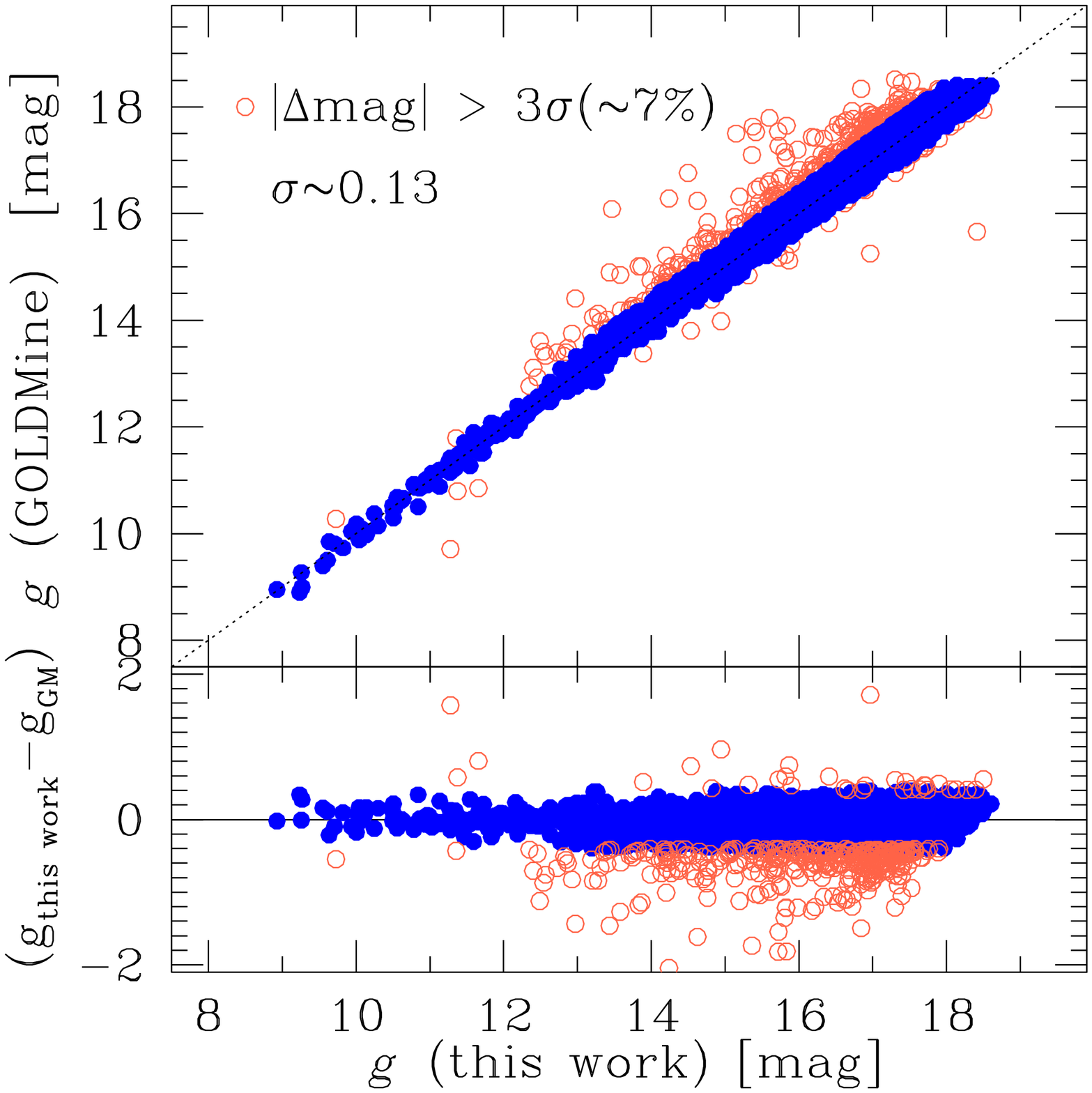} \includegraphics[width= 7.5cm]{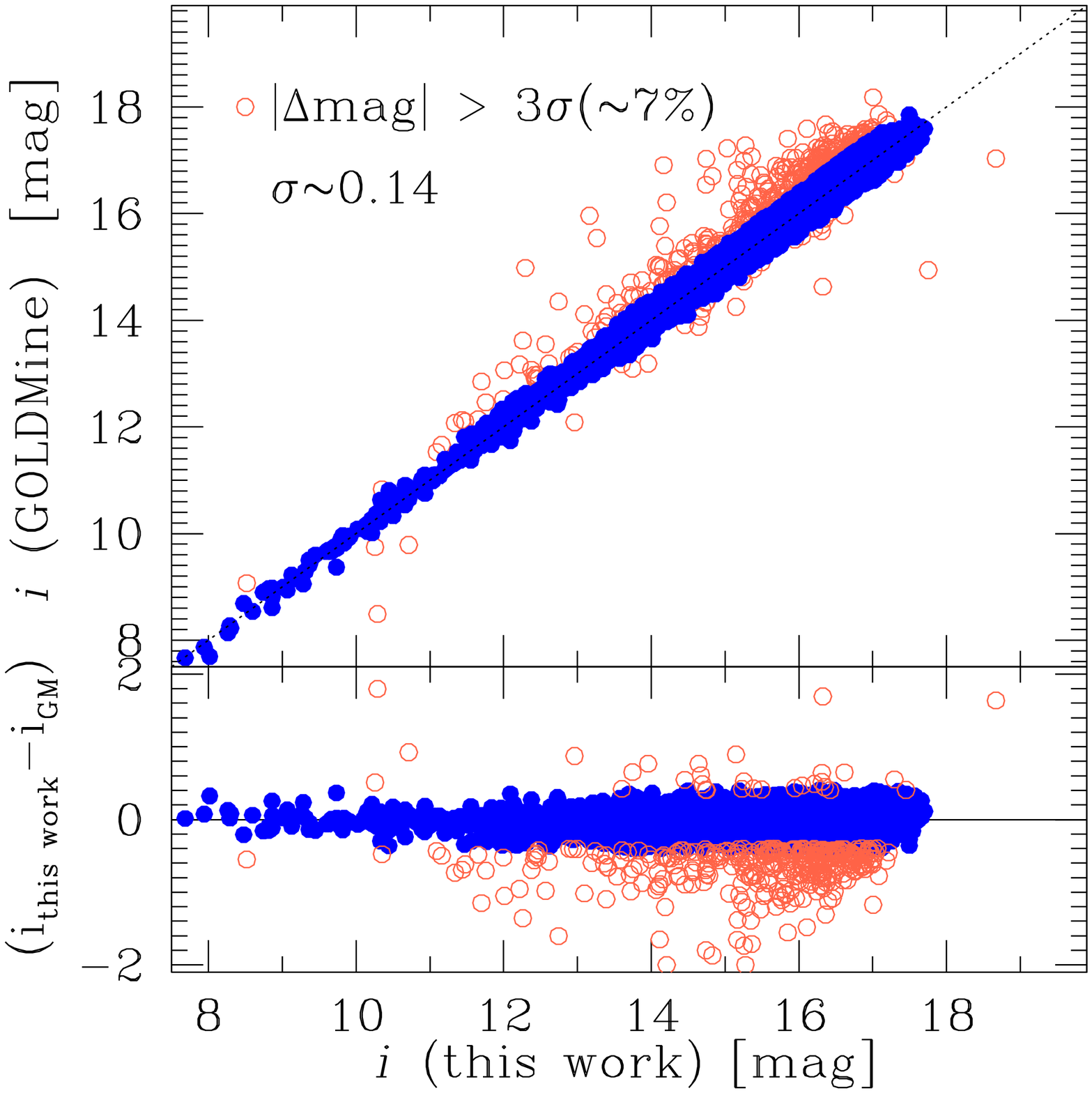}
\includegraphics[width= 7.5cm]{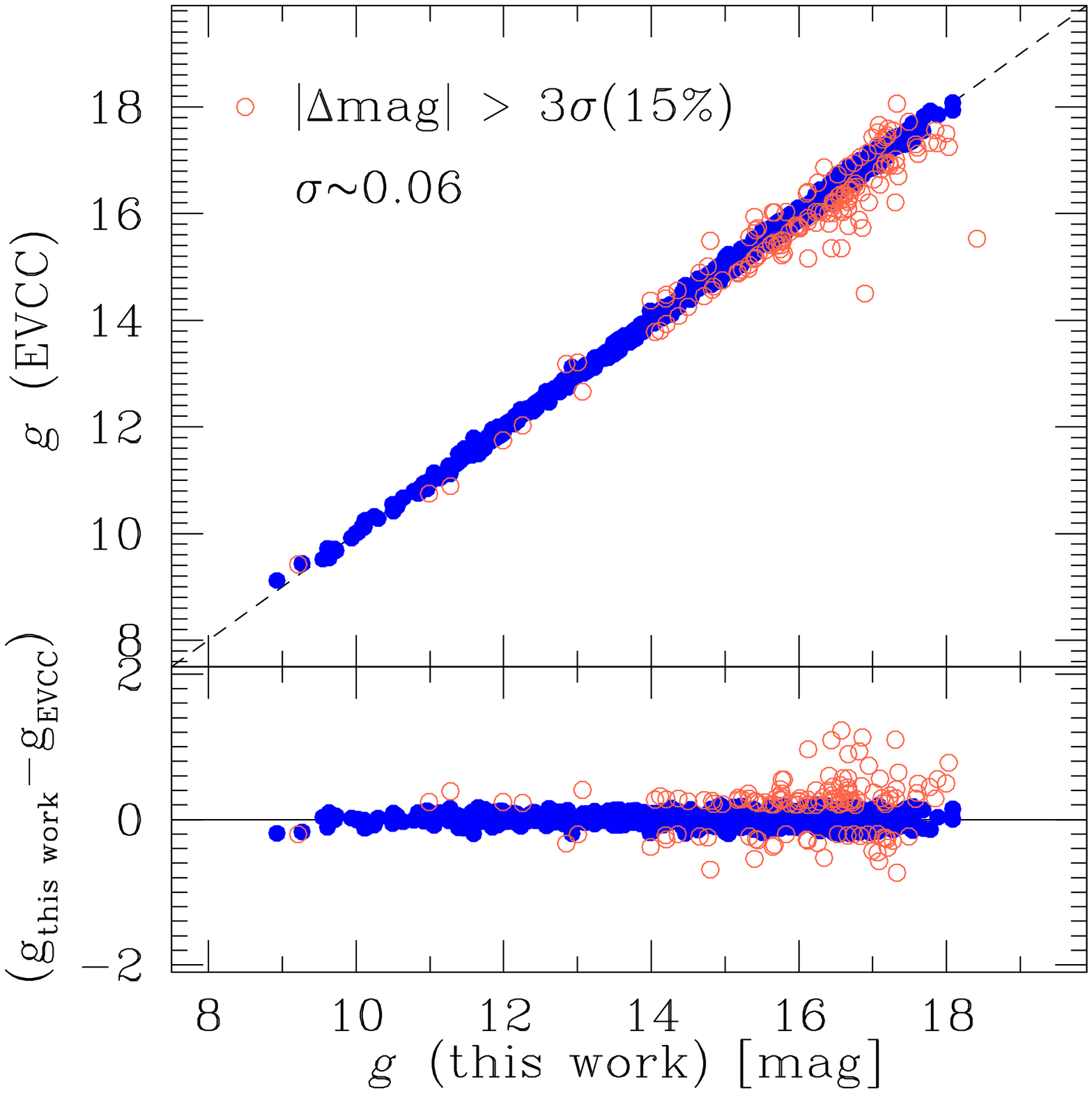} \includegraphics[width= 7.5cm]{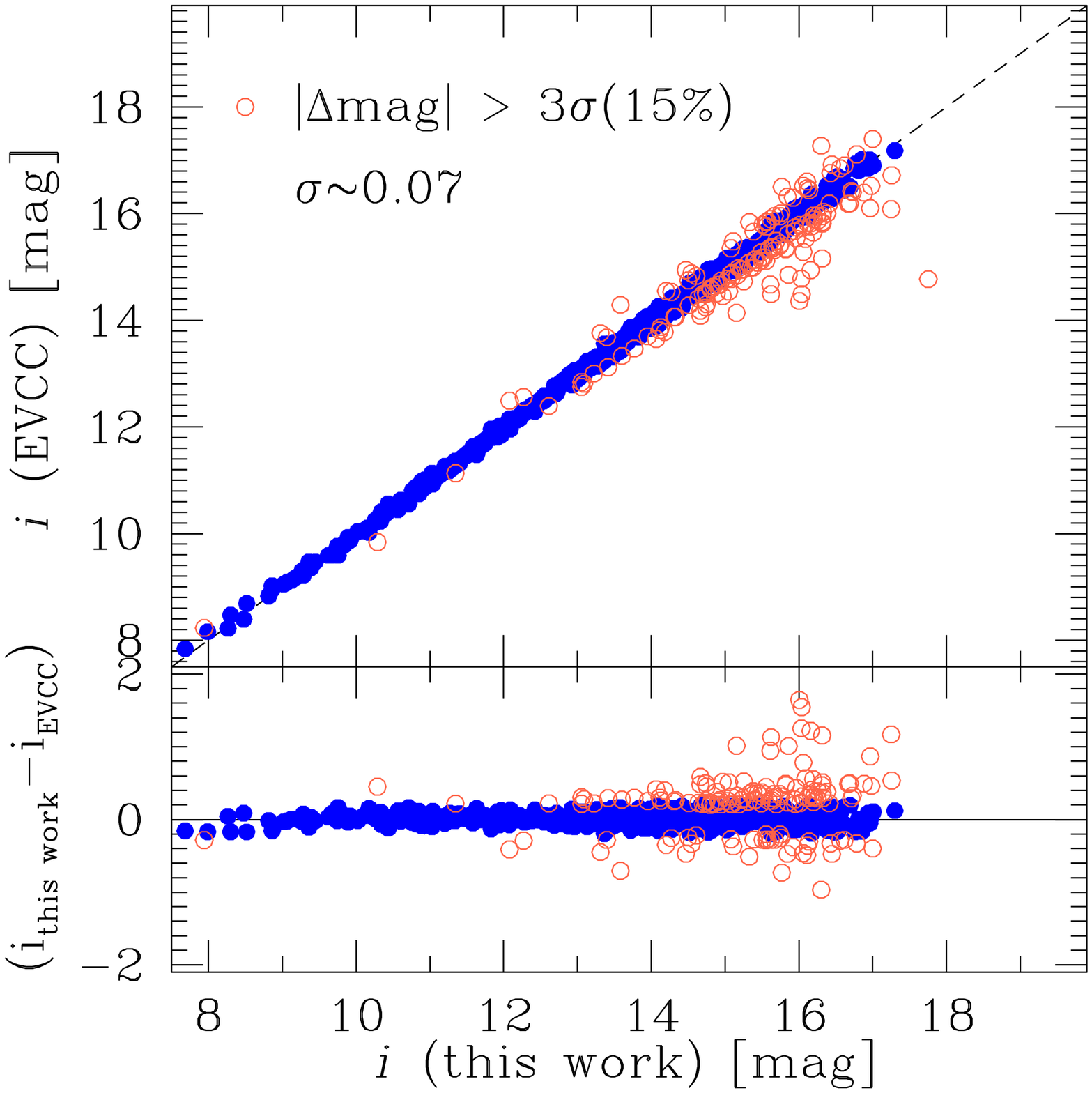}
\caption{
The $g$-band and $i$-band Petrosian magnitudes from this work
  compared to: magnitudes from the SDSS Data Release 10 \citep[][top
    panels]{dr10}, those published in Gavazzi et al. (2012, 2013, middle
  panels), and from the Extended Virgo Cluster Catalog \citep[][bottom
    panels]{Kim+14}. Blue points refer to the data within 3$ \sigma$ from the
  one to one correlation, while the orange points highlight the outliers i.e., galaxies with $\Delta mag > 3\sigma$ to the one-to-one relation  (where sigma is the standard deviation of the
  residual distribution). In each panel we report 
  the sigma of the residual distribution as well as the percentage of outliers. We
  stress that the agreement with \citet{pg12, pg13b, Kim+14} is satisfactory,
  in particular for the brightest objects (42), that are totally
  missing in the SDSS DR10.  For each plot, the bottom panel reports the
  residual magnitudes between the two measurements.}
\label{magGM}
\end{centering}
\end{figure*}

For the 5532 galaxies analyzed with our procedure, we  compute $g$ and $i$ Petrosian magnitudes, 
surface brightness profiles and $(g-i)$ color profiles truncated at 1$\sigma$ of
  the background. Petrosian AB magnitudes and colors are given in Tab.\ref{tab1} and are available via the online 
database GOLDMine \citep{pg03,pg14b}. 
  \begin{figure}
\begin{centering}
\includegraphics[scale=0.45]{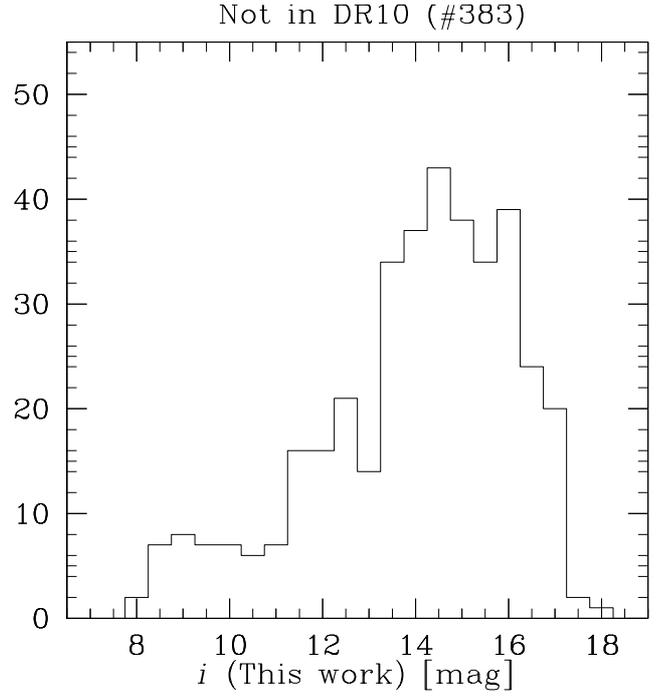}
\caption{Distribution of magnitudes that are missing in the 
SDSS catalog as a function of the $i$-band apparent magnitude from this work. }
\label{mag10}
\end{centering}
\end{figure}
\subsection{Magnitudes}
\label{mags}
In the top panel of Fig \ref{magGM}, the Petrosian magnitudes extracted with the procedure described above are compared 
 to those downloaded from the data release 10 of the SDSS \citep{dr10}. We found values for
5465 objects, while 383 ($\sim 7\%$) objects are missing in the DR10: 274 are from the
Local supercluster and 113 from the Coma supercluster.  
Although a considerable fraction show an appreciable
agreement (residuals have a $\sigma \sim 0.13$ in both the $g-$ and $i-$ band and median value of $-0.05$; $\sim 90 \%$ of the data is in agreement within $0.4$ mag,
i.e. $3\sigma$), 
about $20\%$ of the SDSS magnitudes differs of more than 0.3 mag from our determinations and about
2 \% of the plotted magnitudes show up to 8 mag discrepancy in the SDSS DR10 database values.

One problem can be ascribed to the SDSS determination of the local background, i.e. by cutting all pixels
above a certain sigma level and considering all remaining ones as being part of the background. For large, diffuse, extended objects
this leads to a  wrong local background determination that is used by the pipeline.
The distribution of the residuals  (bottom diagram of the top panels of Fig. \ref{magGM}) saturates at faint magnitudes indicating a systematic effect in 
small and low surface brightness systems
(5th percentile at $\sim -0.75$ while the 95th percentile falls at $\sim 0.05$). 
We checked in the photometric catalog of the DR10 for possible 
causes of such an effect and found that
 the most deviant points are indeed low surface brightness, blue systems that are flagged as NOPETRO by the SDSS pipeline. This means that their Petrosian radius could not 
be measured by the pipeline in the $r-$ band \citep[because of their low SN,][]{pipeline, pipeline2} and it is set to the PSF FWHM, thus  deriving PSF magnitudes
for objects which are indeed extended.
The other (less extreme) deviant points are again irregularly shaped, blue galaxies all reporting photometric flags indicating problems either with the raw data, the image or 
the evaluation of the Petrosian quantities \citep{pipeline, pipeline2} such as: DEBLENDED\_AS\_PSF (deblending problems), DEBLEND\_NOPEAK, MANYPETRO,
\footnote{In irregular low surface brightness systems there may be more than one Petrosian radius, \citet{pipeline}} etc.
In these cases, the SDSS pipeline underestimates the Petrosian radius and hence the aperture for the photometry.
Henceforth we stress that our procedure is not affected by this problem 
because it measures the Petrosian radius in the white image which has an improved SN ratio.
 This allows us to derive reliable magnitudes even in low surface brightness systems and, in general, systematically
measure a larger Petrosian radius (hence Petrosian flux) if compared to the SDSS $r$-band Petrosian radius. 
Nevertheless we stress that, the fact that the Petrosian flux is taken within two Petrosian radii ensures that both apertures gather substantially the same Petrosian flux 
for the vast majority of the sample \citep{kron}, i.e., the total flux.
Moreover, DR10 lacks almost completely the brightest and most extended galaxies
($g \lsim 11$, $i \lsim 10$), as demonstrated by the distribution of the 383 objects that are not included in
the SDSS database shown in Fig. \ref{mag10}.

In fact, considering both discrepant and missing magnitudes, $\sim 9\%$ of the SDSS data are
either unreliable or missing. In particular, in the Local supercluster, the
percentage of missed galaxies of the DR10 is around 26\% and, if we consider objects below 11
magnitudes, it reaches a dramatic 95\%, reflecting the difficulties of the
SDSS pipeline when dealing with nearby extended objects.  
Moreover, the few (12) bright galaxies ($g<11$ mag) included in the DR10 deviates more than $3\sigma$ from 
the one-to-one correlation in Fig.\ref{magGM}.  On the other hand,
in the Coma supercluster, the percentage of missing and unreliable galaxies
drops to 3\%  thanks to the smaller angular size of the objects.

\begin{figure}
\begin{centering}
\includegraphics[scale=0.38]{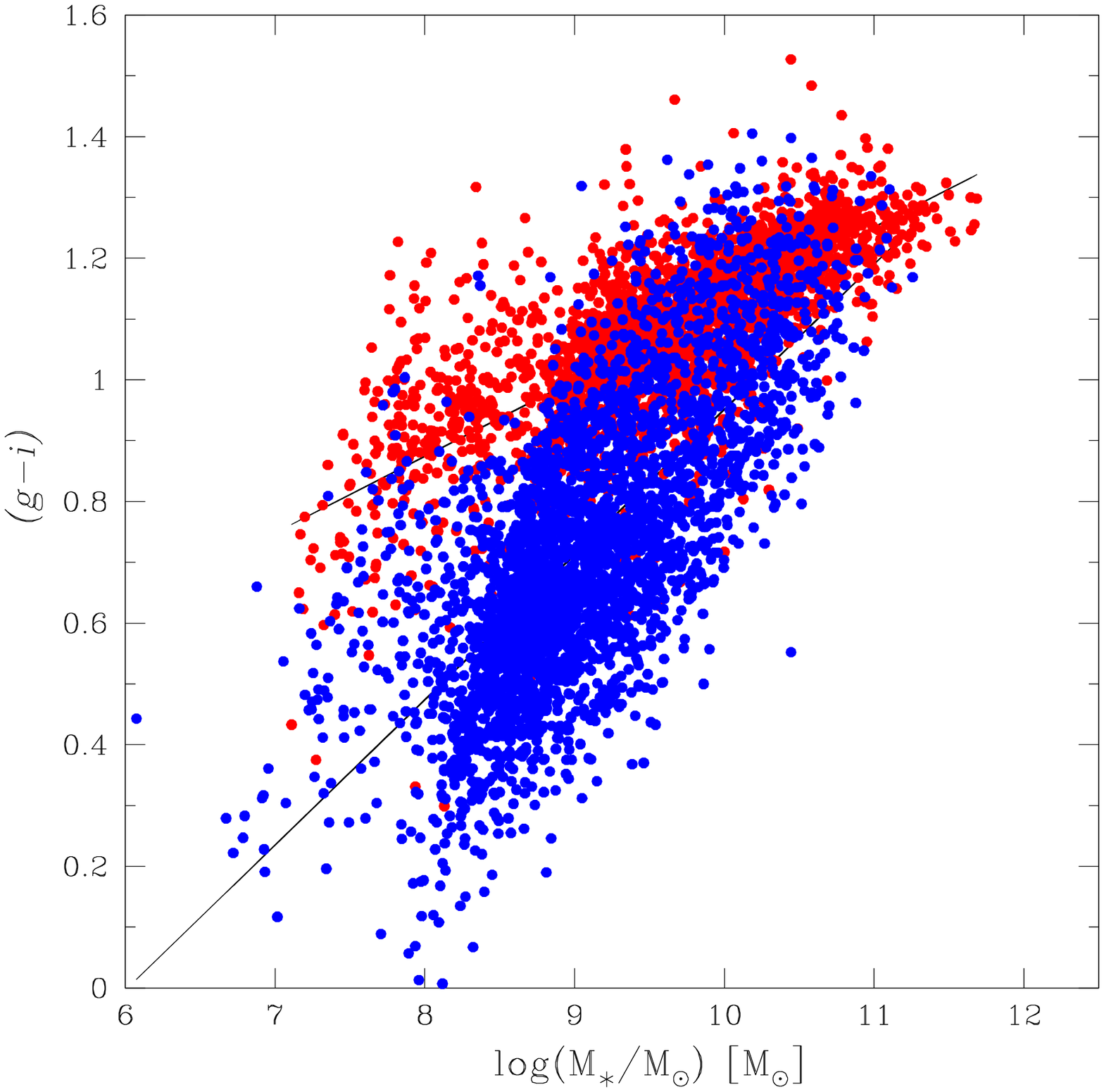}
\includegraphics[scale=0.38]{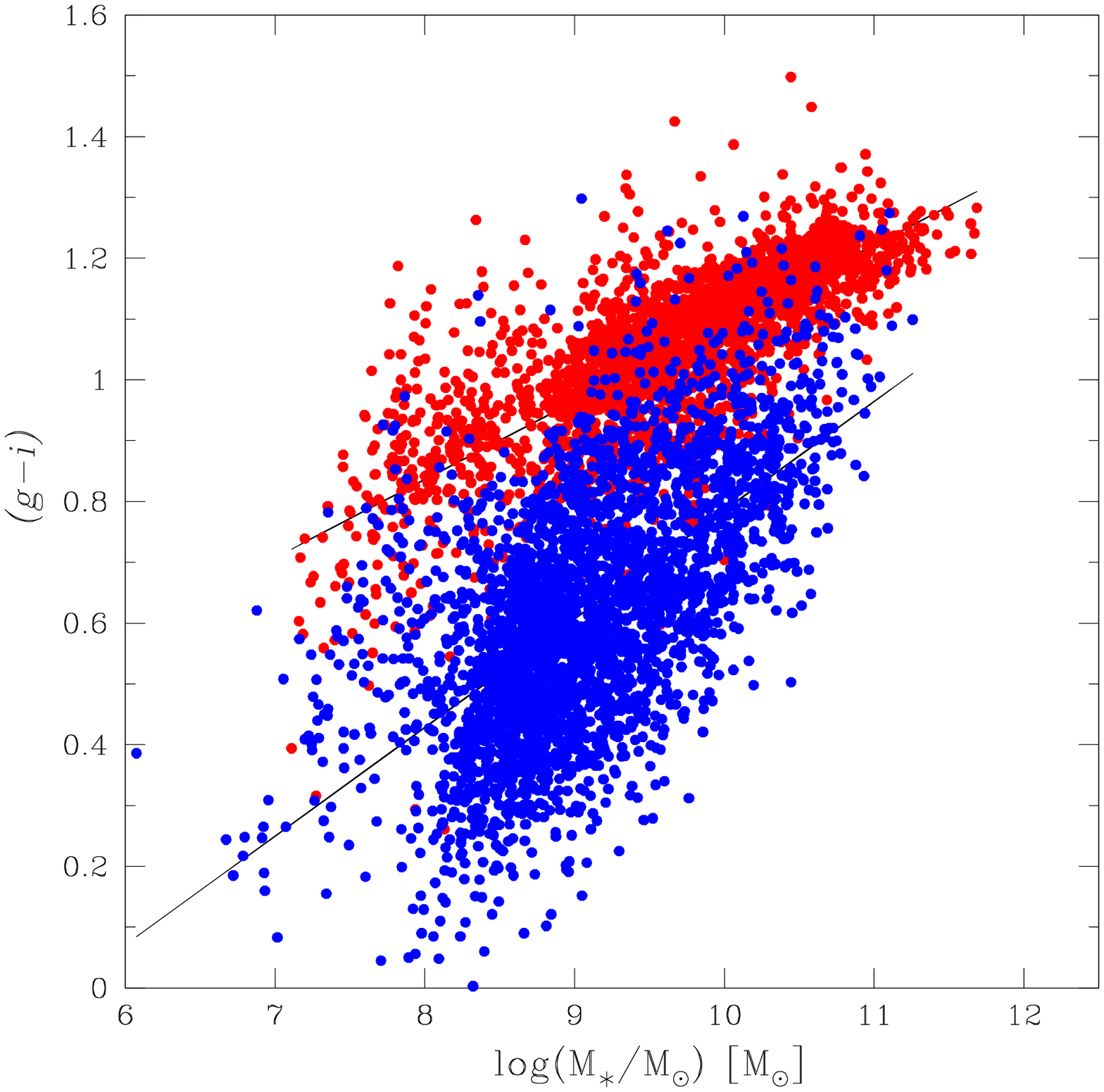}
\caption{Top: Color magnitude diagram of the observed colors displaying the well known bimodal
 distribution for late type galaxies (blue cloud, blue dots)
and early type galaxies (red sequence, red dots). Bottom: the color magnitude diagram after the
inclination correction has been applied. The red sequence and the blue
cloud are well separated at all masses.}
\label{colmagss}
\end{centering}
\end{figure}

To overcome the above discrepancies, about 400 galaxies of the Local
supercluster have been manually measured using IRAF/QPHOT by \citet{pg12}.
Moreover, for nearly 5000 galaxies in the Coma Supercluster the magnitudes
were derived from the SDSS DR7 by \citet{pg13b}.  We compare our
automatically extracted magnitudes with those from \citet{pg12,pg13b} \citep[which are publicly available via the online database GOLDMine][]{pg13c, pg14b}
obtaining a satisfactory agreement (median difference of $\sim -0.025$ mag), as shown in Fig. \ref{magGM} (middle
panels), especially below 11 mag.  At the faint end there is less agreement,
as $\sim 15 \%$ of galaxies show a residual between the two measurements that
exceeds 0.3 mag. 
The 5th percentile of the residuals distribution 
is $\sim -0.45$ mag while the 95th percentiles is $\sim 0.2$ mag, indicating that the systematic effect 
at low surface brightness affecting the DR10, affects the Gavazzi et al. measurements too.
In fact, despite Gavazzi et al. checked the reliability of the magnitudes taken from the SDSS database of the magnitudes of the DR7, these were not 
re-calculated and therefore suffer similar problems as the DR10 data. As a matter of fact, 
the most discrepant objects are once again mainly blue and low surface brightness and are flagged in the SDSS photometry as problematic data. 
Hence, we believe that the observed discrepancies are again to be attributed to a bad evaluation of the Petrosian radius by the SDSS pipeline among some 
irregular and low surface brightness systems.\\ 
More recently, \citet{Kim+14} released the new Extended Virgo Cluster Catalog
(EVCC) which covers an area 5.2 times larger than that of the VCC catalog
\citep{vcc}. It  includes 676 galaxies that were not included in the original VCC
catalog. Similarly to our work, the EVCC is based on SDSS DR7 
images, but it  includes all the five SDSS bands $u$ $g$ $r$ $i$ $z$
determined using Source Extractor with parameters that are 
tailored by inspection of the individual galaxies.\\ We found 844 galaxies in
common with the EVCC and we compared their $g$ and $i$ magnitudes with our
results.  The resulting correlation is plotted in Fig.\ref{magGM} (bottom
panels) and shows a more than satisfactory agreement  (median residual of -0.009 mag, $\sigma\sim 0.07$) between the two
measurements.  In particular, we remark the good agreement reached below 14
mag indicating that, unlike the SDSS photometric pipeline,  neither the shredding
problem nor the aperture radius measurement affect the results of both semi-automated methods. Moreover, 85\% of the galaxies differs by less than 0.2 mag between 
the two studies. 
The \citet{Kim+14} magnitudes are measured by Source Extractor within an aperture with radius equal to $k$ times the Kron radius \citep{kron} of the galaxy, 
where $k$ is the Kron factor and is set to $k=2.5$. This value is chosen because is expected to recover more than $94\%$ of the total flux \citep{sex96}.
Therefore the agreement further demonstrate that the aperture radius set to two times the Petrosian radius evaluated in the white image recovers efficiently 
the total flux of the galaxy. 
Despite the overall excellent agreement, 6\% of galaxies above mag $\sim 15$ display a
difference between our and their measurements that exceeds 0.3 mag ($\sim5\times \sigma$) and, 
in two cases, the two determinations differs of $\sim 3$ mag.
The vast majority of these discrepant objects appear significantly fainter in our analysis and are low surface brightness systems that lie very near to bright sources 
(e.g., stars, interacting galaxies etc.). We re-measured these magnitudes with IRAF/QPHOT and concluded that 
some contaminating light has likely not been fully masked by \citet{Kim+14}. 

\section{The color-magnitude}
\label{colmags}

After computing the observed magnitudes as described in the previous sections,
we correct them for Galactic extinction, following the 100 micron based
reddening map of \citet{dust98}, re-calibrated by \citet{dust11}.
Furthermore, colors of late-type galaxies are corrected for internal
extinction using the empirical transformation \citep{pg13b}
\begin{equation}
(g-i)_{0}=(g-i)_{mw}-0.17([1-\cos(incl)][\log(\frac{M_{\star}}{{\rm M_{\odot}}})-8.17]),
\end{equation} 
where $(g-i)_{mw}$ is the color corrected for Milky Way Galactic extinction, $M_{\star}$ is the mass computed with eq. \ref{masseq} 
using the uncorrected color and $incl$ is the inclination of the galaxy, computed following \citet{incl}.\\
Stellar masses are derived from the $i$ magnitudes and the inclination 
corrected color $(g-i)_{0}$, assuming a Chabrier IMF via the mass vs $i$-band luminosity relation
\begin{equation}
\log\left(\frac{M_{\star}}{{\rm M_{\odot}}}\right)=-0.96+1.13(g-i)_{0}+\log\left(\frac{L_{i}}{{\rm L_{\odot}}}\right)
\label{masseq}
\end{equation}
published by \citet{zibi09}, where L$_{i}$ 
is the $i$-band luminosity of the galaxy in solar units.\\

Fig. \ref{colmagss} (top panel) shows the observed color-mass diagram of the Local and Coma superclusters before the inclination correction is applied. The sharp oblique density contrast in the figure occurring
around $10^8 $M$_\odot$ is caused by the SDSS selection ($r<17.7$) at the different distances of the Local and Coma superclusters (respectively $\sim 17$ Mpc ad $\sim 95$ Mpc) and
by the mild dependence on galaxy color of the mass-luminosity relation. 
The Local supercluster is indeed less populated due to the lack of sampled volume and is undersampled at high masses. 
Nevertheless, owing to its proximity, the SDSS selection allows us to include in the analysis
galaxies down to masses as low as $\sim 10^{6.5}$M$_\odot$. On the contrary,  the selection restricts the Coma color-magnitude relation  to  $\gsim 10^{8.5}$M$_\odot$
making the blue cloud relation appear steeper than the local one (see Tab. \ref{slopes}). 
\begin{table}[t]
\caption{Slopes of the linear fits $(g-i)=a+b\times\lg(M_*)$ of the blue cloud and red sequence for the Local supercluster (Lsc), the Coma supercluster (Csc) and the whole sample (All).}
\centering
\begin{tabular}{c c c }
\hline\hline
    Sample  & $b_{LTG}$    &	  $b_{ETG}$		  \\
            &		   &				  \\
\hline   
     Lsc    & $ 0.148 \pm   0.007$ & $ 0.121 \pm 0.005  $ \\
     Csc    & $ 0.214 \pm   0.005$ & $ 0.150 \pm 0.004  $ \\
     All    & $ 0.184\pm    0.004$ & $ 0.131 \pm 0.002  $ \\
 \hline
\end{tabular}
\label{slopes}
\end{table}

Early-type galaxies  (dE,dS0, E, S0 and S0a; red dots) form the red sequence 
while late-type galaxies  (from Sa to Irr, blue dots) follow the blue cloud that overlaps with the red 
sequence at high masses, as highlighted with early SDSS data by \citet{strat01},\citet{hogg04}.   
The morphological cut is based on the morphological classification available in GOLDMine \citep{pg03,pg14b}.
After the internal extinction 
correction is applied, as in Fig. \ref{colmagss} (Bottom panel), the color magnitude preserves its 
bimodality, with less overlap at  high masses.  At this point a note of caution is required: the selection bias that affects the more populated Coma sample prevents us from reliably 
extract a general slope. Therefore we constraint our fit to the deeper data of the local Universe and find that
the slope of the blue cloud remains steeper than that of 
the red sequence even after the corrections. As a matter of fact, the local blue cloud best fit  changes from
$(g-i)_{LTG}=0.18\times \log(M_*)-0.87$ in the uncorrected plot to
\begin{equation}  
(g-i)_{LTG;0}=0.15\times \log(M_*)-0.64 
\end{equation}
when the correction is applied, where $(g-i)_{LTG}$ and $(g-i)_{LTG;0}$ are respectively the blue cloud uncorrected and corrected color and the error over the slope is as low as $\sim3\times10^{-3}$ in both
relations.
 As expected, since no inclination correction acts on ETGs, the red sequence fit remains consistent in both relations: \\
 $(g-i)_{ETG}=0.13\times \log(M_*) -0.12$.
\begin{figure*}
\begin{centering}
\includegraphics[width= 17cm]{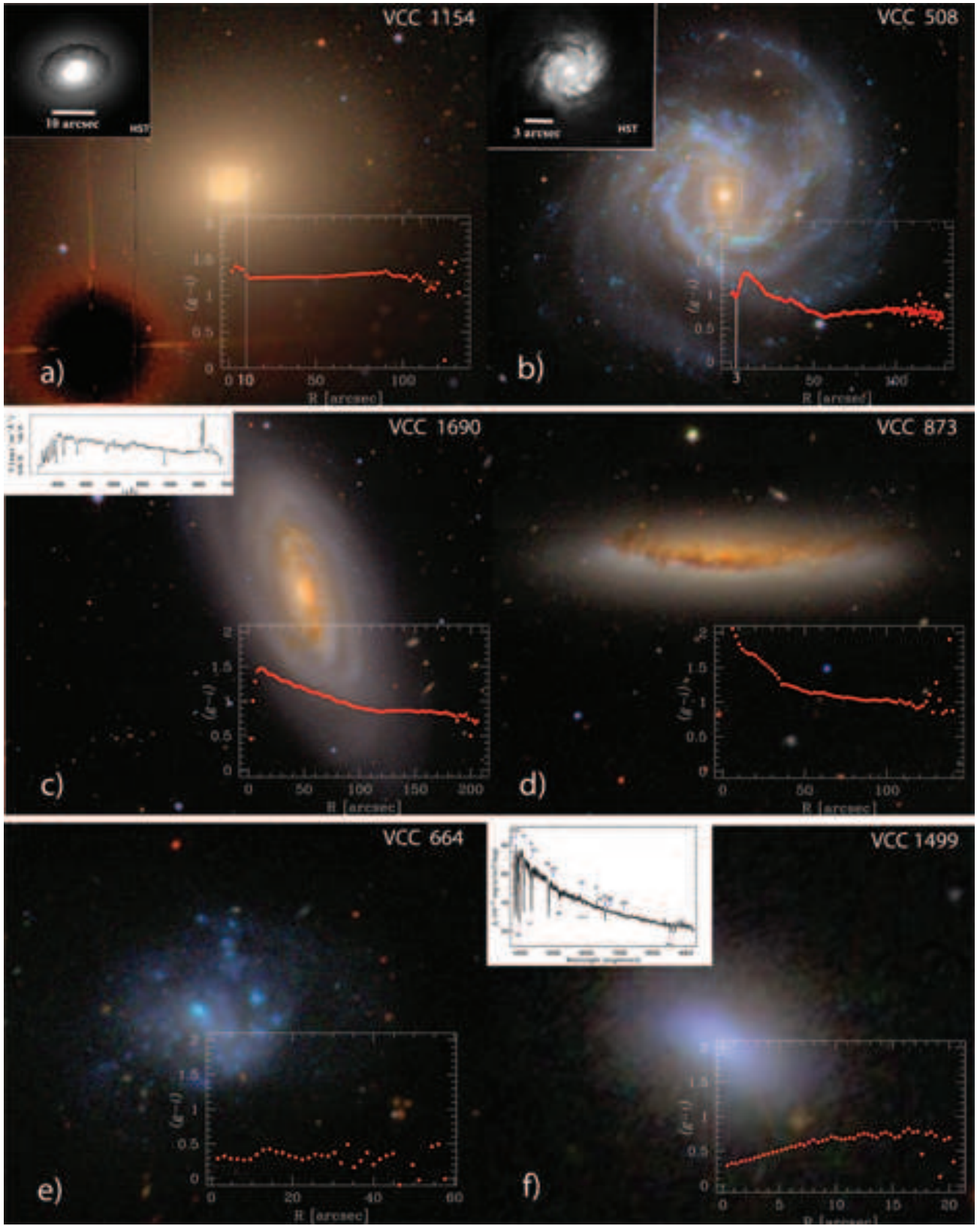}
\caption{RGB SDSS images and (g-i) extracted color profiles of six galaxies belonging to the Local supercluster sample: 
(a) VCC 1154 is an S0 galaxy with a circumnuclear dusty disk of $\sim 800$ pc as  shown by an archival HST image. 
The presence of these structures is easily spotted by a clear deviation towards the red on circumnuclear scale 
in the (g-i) profile  at $3"$ as indicated by the vertical white line. (b) The barred spiral galaxy VCC 508. Its circumnuclear star-forming disk (highlighted in the inset)
produces a deviation towards blue in the color profile  at $\sim10"$ as indicated by the vertical white line. (c) VCC-1690 (Boselli et al. 2016), spiral galaxy with a strong blue AGN nuclear 
(g-i) color. Its nuclear spectra (shown in the inset) exhibit the key signatures of post starburst galaxies. 
(d) Late-type galaxy VCC-873. The AGN activity is strongly obscured by dust and its color profile  reaches extreme values.
(e) Irregular galaxy VCC 664 and its blue, nearly flat color profile. (f) The low mass galaxy together with its PSB-like 
nuclear spectrum. Its central emission strongly deviates towards bluer colors with respect of 
the outermost region of the galaxy.}
\label{profils}
\end{centering}
\end{figure*}
\section{Radial color profiles}
\begin{figure} 
\begin{centering}
\includegraphics[scale=0.2]{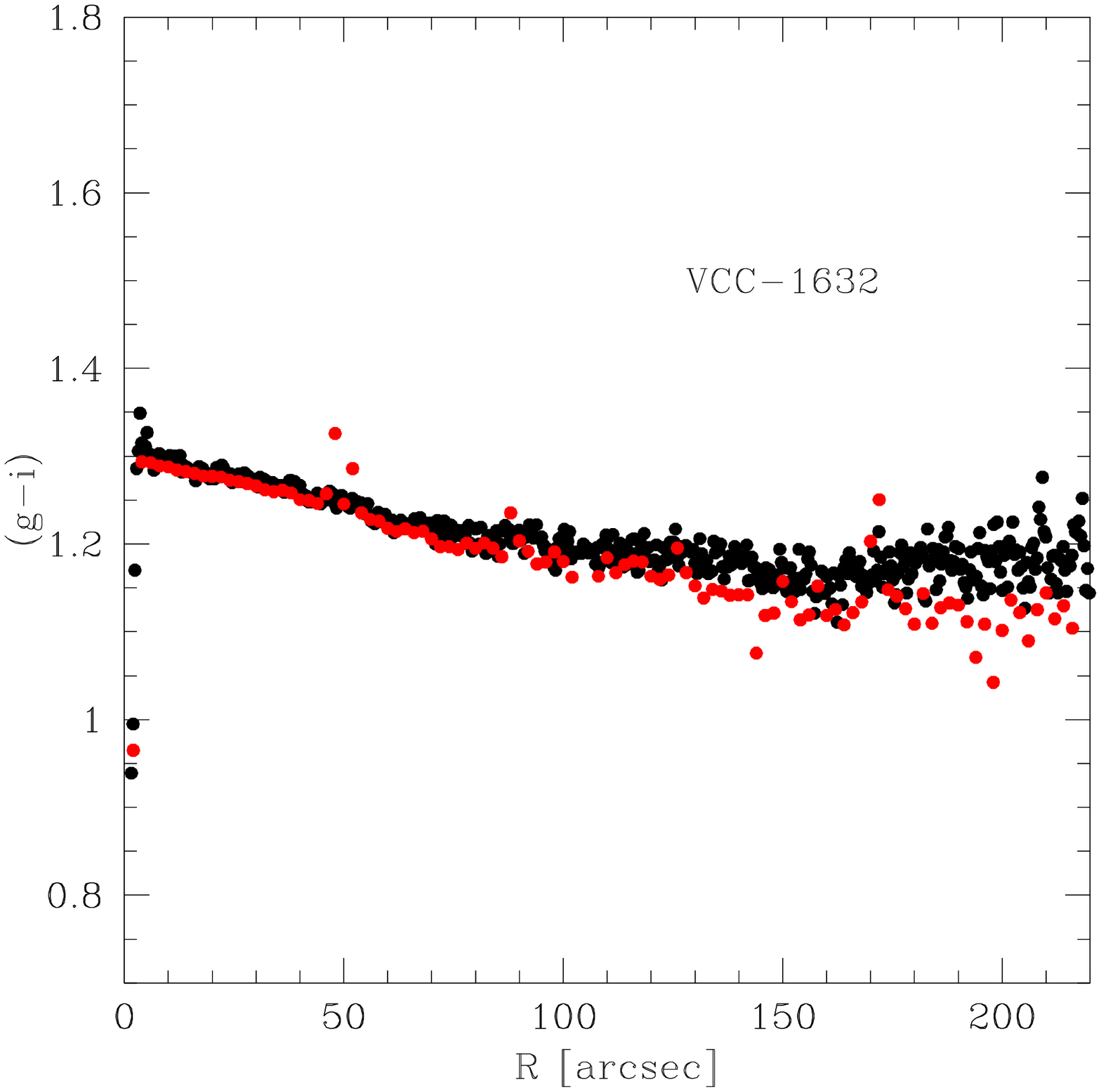}\includegraphics[scale=0.2]{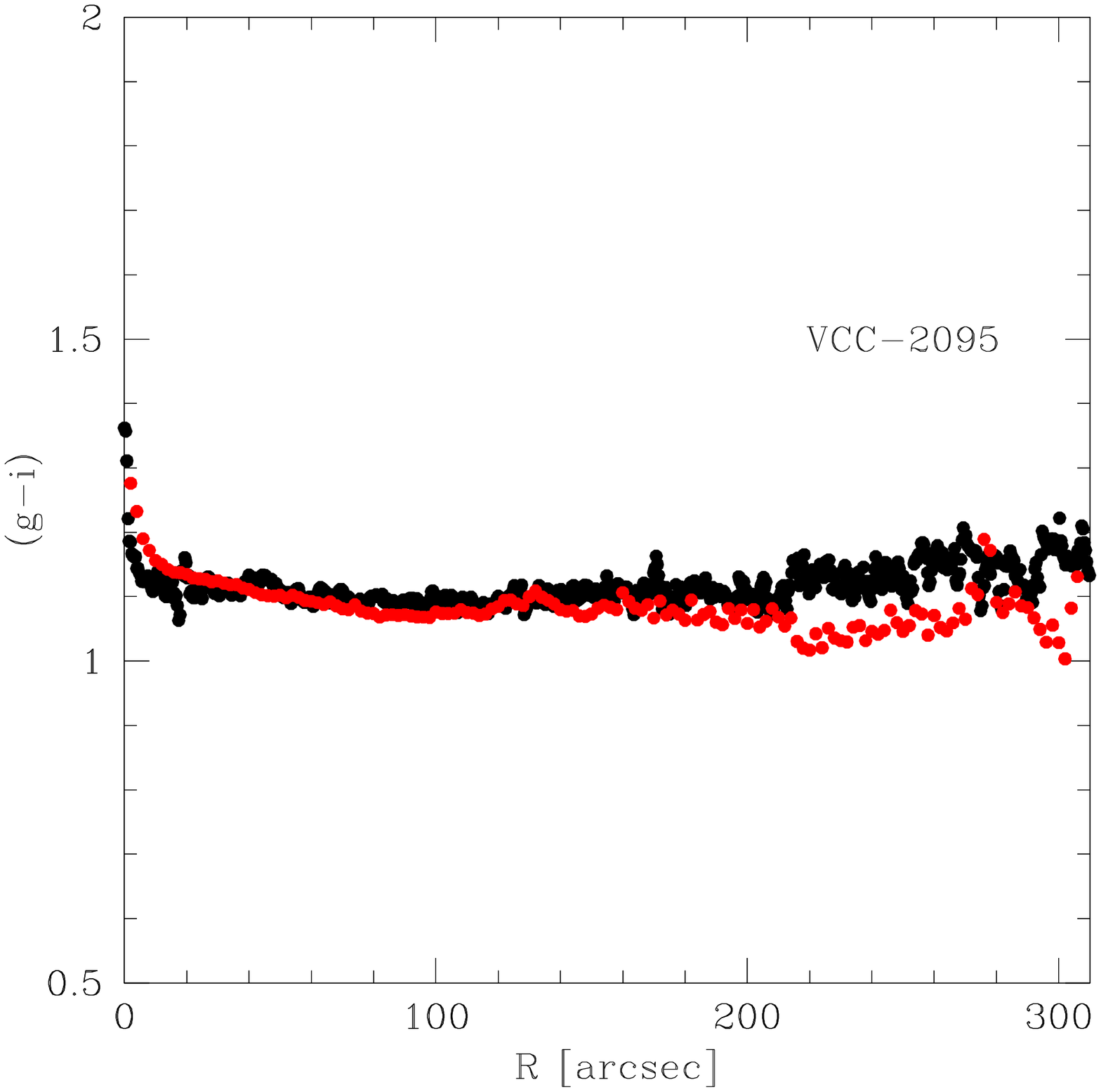}
\includegraphics[scale=0.2]{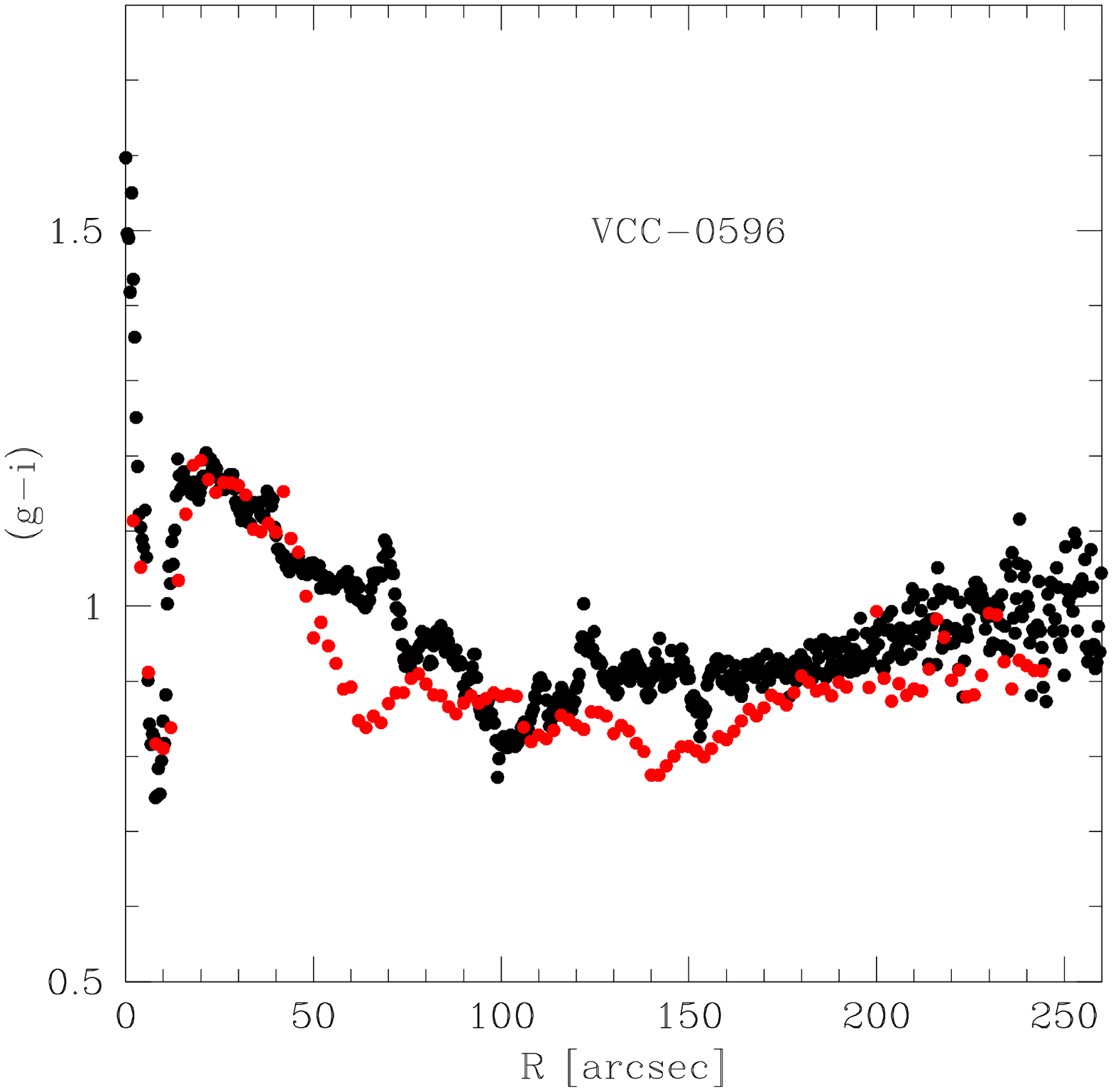}\includegraphics[scale=0.2]{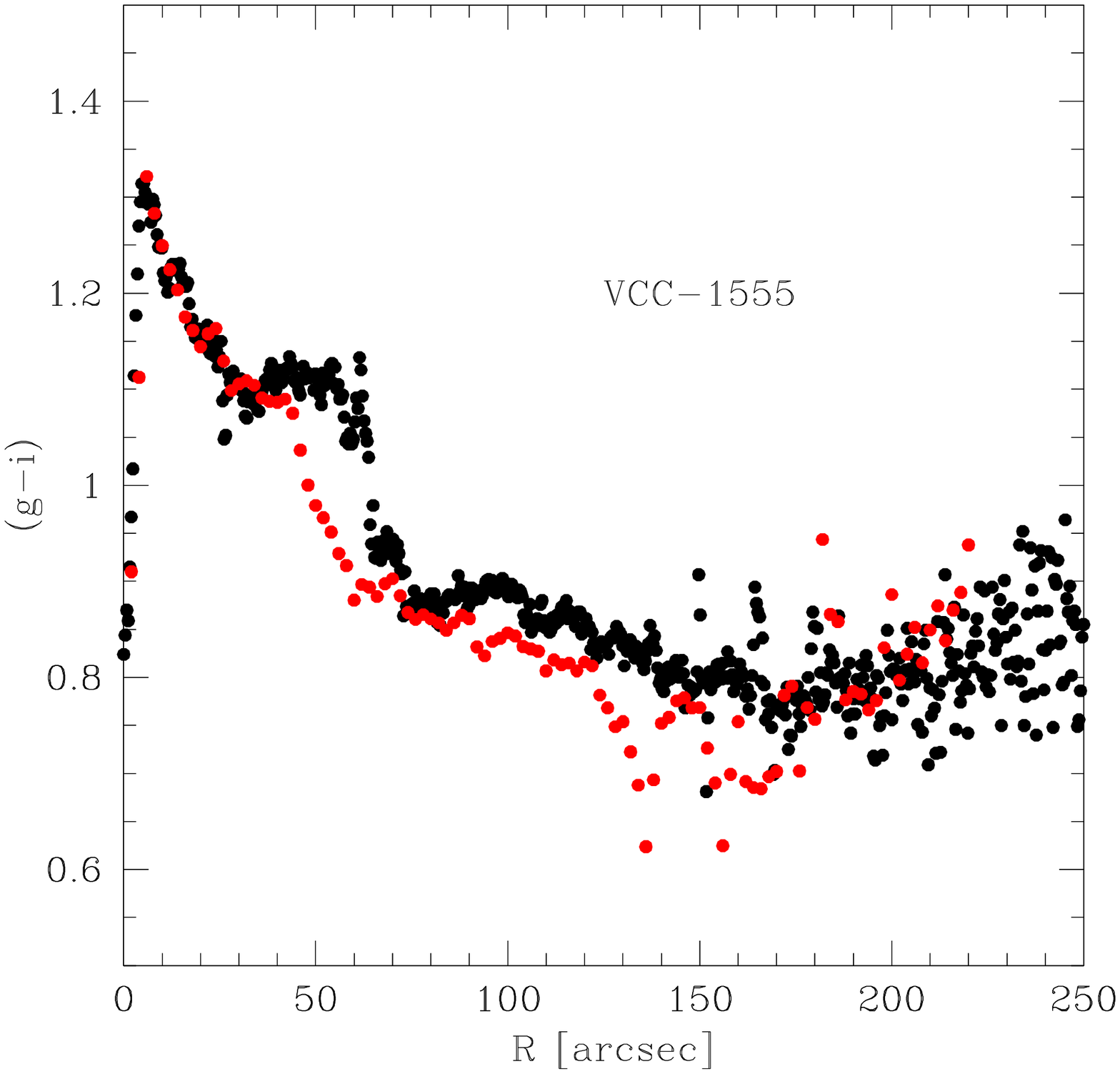}
\includegraphics[scale=0.2]{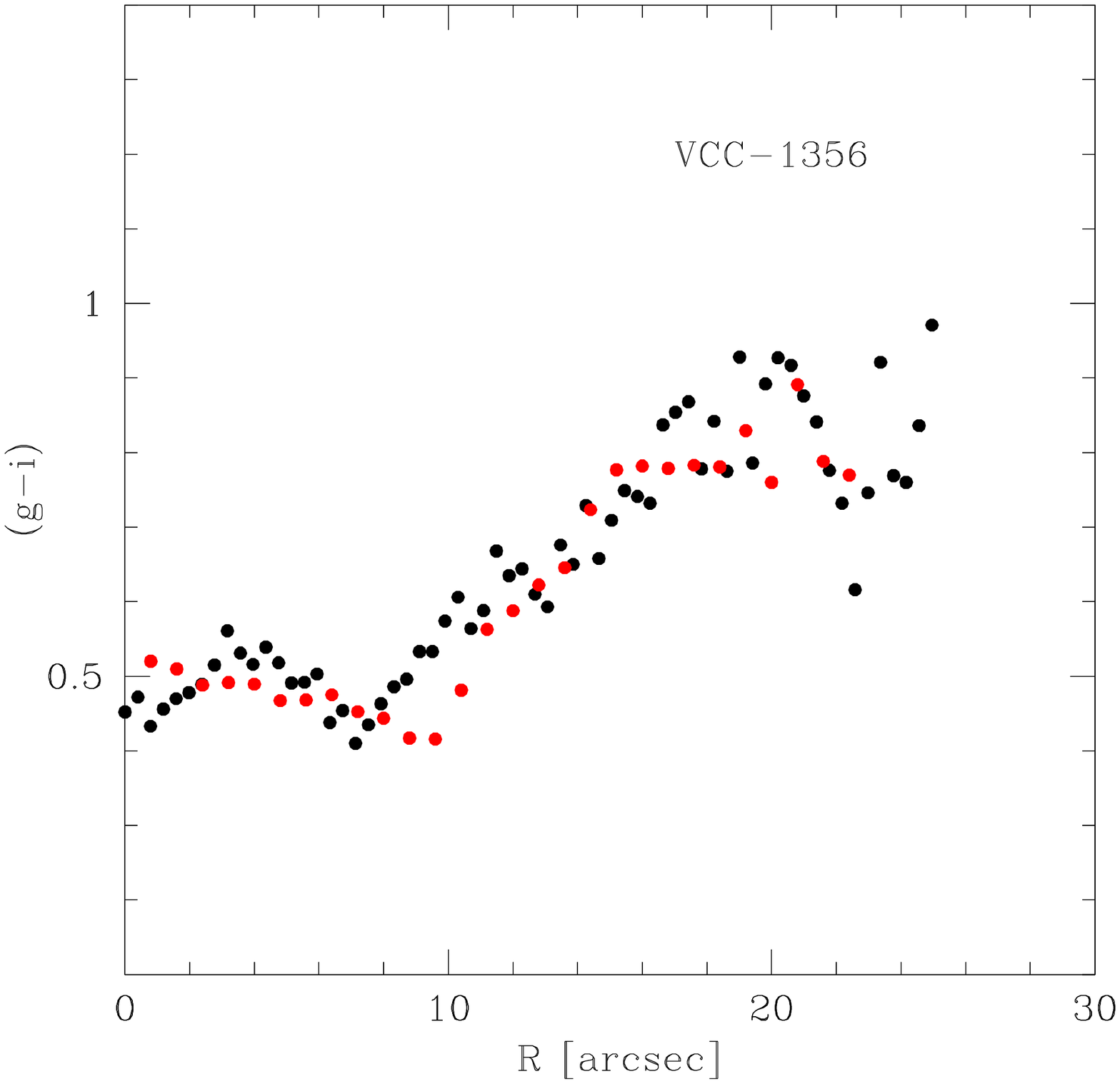}\includegraphics[scale=0.2]{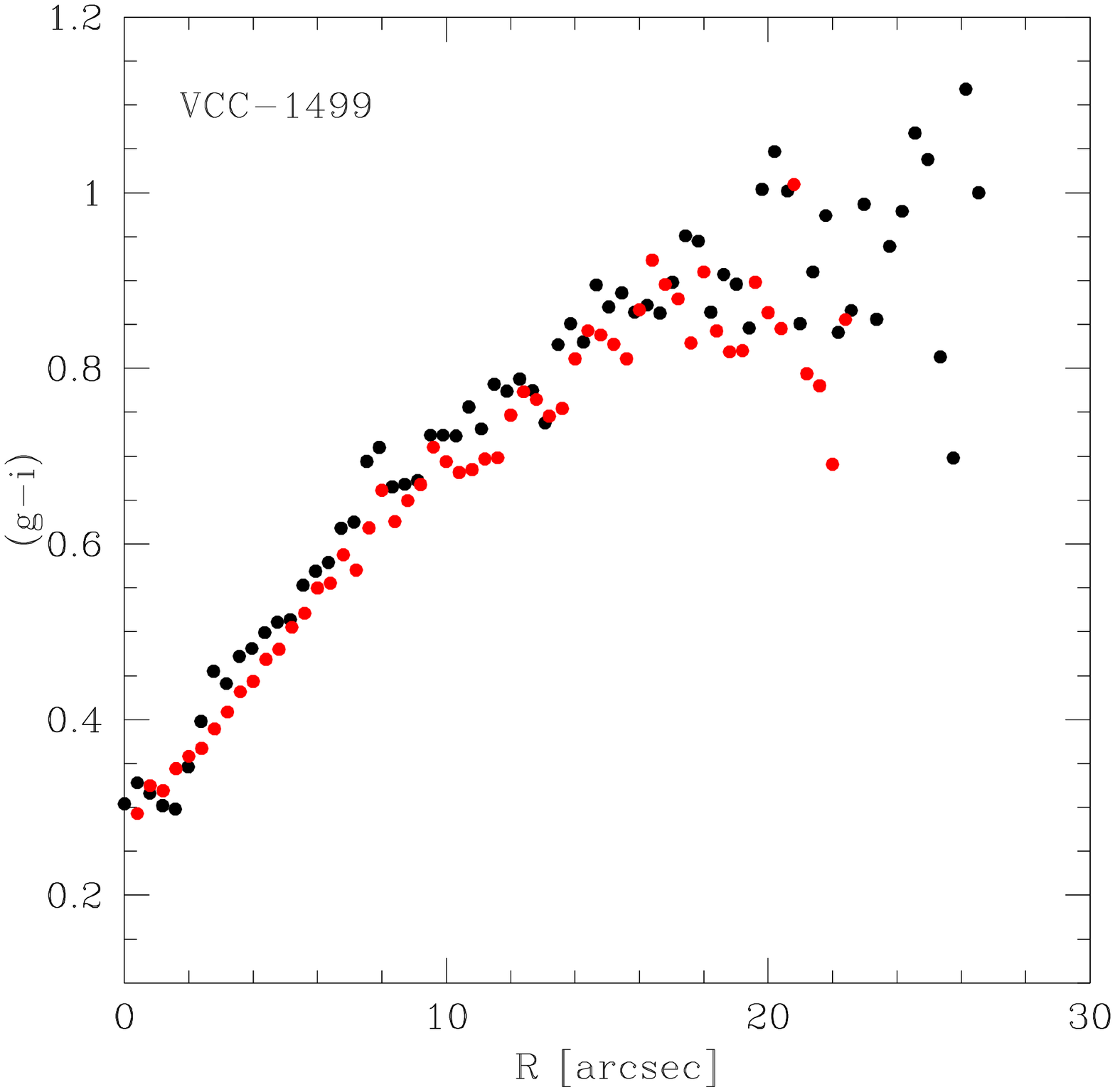}
\caption{Comparison of our color profiles (red dots) with the ones derived from McDonald et al. (2011) (black dots) for six galaxies matching the two samples
of different types. From top left to right we plot
VCC 1632, VCC 2095, VCC 596, that are respectively an E,S0,Sc (from NED). In the second row we plot VCC 1555, VCC 1356, 1499, respectively an Sc an Sm and a dE. There is an overall good agreement but the profiles 
display some differences especially in late type galaxies. We ascribe these effects to the difference in the isophotal fitting technique and the one adopted in this work to 
extract the profiles. McDonald et al. (2011) are indeed sensitive to twists of the isophotes occurring in the presence of spiral arms, bars and even in bright HII regions in irregular galaxies.}
\label{courteau}
\end{centering}
\end{figure}
\label{colori}
\begin{figure*}
\begin{centering}
\includegraphics[scale=0.9]{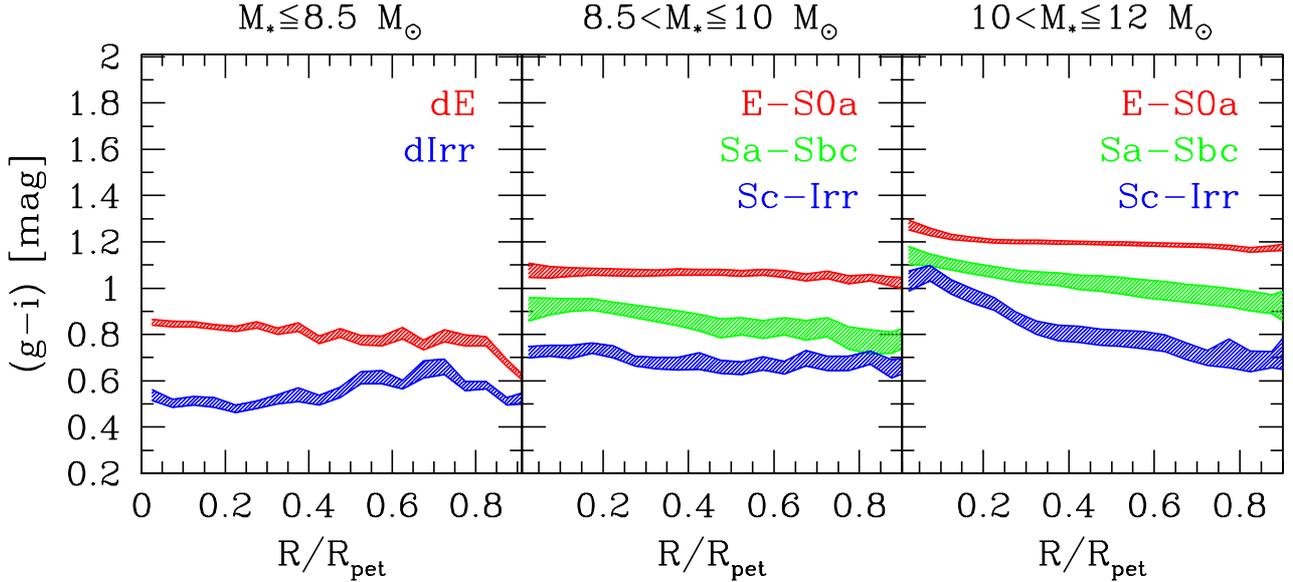}
\caption{Template color profiles in different bins of mass and morphology.
  Left panel: low mass bin (M$\le 10^{8.5}$M$_{\odot}$) where red stands for
  dwarf ellipticals (dE) and blue for dwarf irregulars (dIrr). Central panel:
  intermediate mass bin ($10^{8.5}$M$_{\odot}<$M$\le 10^{10}$M$_{\odot}$)
  where red dots are for the ETGs (from E to S0a) template, green dots for
  Spirals (from Sa to Sbc) and blue dots for Sc$-$Irr LTGs. Right panel: high
  mass bin (M$>10^{10}$M$_{\odot}$) where the color code is the same as the
  intermediate mass bin. In both the high-mass and the intermediate-mass bins,
  we did not take into account galaxies with $incl > 60^o$.}
\label{tempmass}
\end{centering}
\end{figure*}
For each object of the sample, our procedure yields a  radial color
profile. In this section, we discuss the color profiles extracted 
by taking a closer look to some prototypical cases. Then we describe more quantitatively the general
properties of color profiles along the Hubble sequence creating a set of radial color profile templates 
of different morphological classes in different mass bins.\\
In general, the color is a good tracer of the specific SFR: regions
actively forming stars have bluer colors than regions of quenched SFR and 
profile shapes generally correlate with the Hubble type. Early type galaxies (e.g., Es and S0s) 
typically have almost flat and red profiles,  common among the red and dead objects \citep{tamura03,wu05}. 
 Differently from the ETGs, irregular galaxies are characterized by an almost flat blue profile 
reflecting their lack of dust and their ongoing SF activity at all radii (Fig. \ref{profils}, bottom row).
Finally spiral galaxies have a composite profile that is blue at  large radii and becomes redder only toward the 
center in correspondence with structures such as bulges or bars \citep[e.g., VCC-1690 in Fig. \ref{profils}, see also ][for a detailed discussion]{pg15},  consistent with
the conclusion drawn by \citet{fox13} who compared the distribution of the star formation from H$\alpha$ imaging flux with the stellar continuum.\\
Strong deviations from the typical color profiles can be induced by strong dust absorption, as 
 in the case of the highly inclined galaxy  VCC-873 shown in Fig. \ref{profils}. 
The strongest deviations are observable in the nuclear regions of galaxies, where, for example, nuclear dusty disks 
are detectable as clear deviations toward a redder color (see, e.g., VCC-1154 in Fig. \ref{profils}, {\it top-left}).
On the other hand, unobscured HII regions, dominated by young stars,
can cause strong deviations towards the blue in the color profiles. 
The barred spiral galaxy VCC-508 (Fig. \ref{profils}, {\it top-right}) is the perfect posterchild of this class of objects, 
with a color profile deviating  towards the blue on a scale of 3 arcseconds ($\sim 300$ pc), in correspondence with  
a star-forming ring (vividly shown in the HST archival image),  and possibly in correspondence of the bar Inner
Lindblad Resonance \citep[ILR,][]{K&K04, comer10, font14}. Similar nuclear blue spikes can be associated to unobscured AGNs 
and post starburst (PSB) nuclei. An example of galaxy hosting a PSB nucleus is VCC-1499 (shown in Fig.~\ref{profils}), a member
of the dEs of the Virgo cluster that display blue nuclei \citep{lisk06}.
To better test the good quality of our profiles in Fig. \ref{courteau} we show the comparison between color profiles extracted in this work with the ones derived from the 
surface brightness profiles published by \citet{mcdonald11} for six galaxies of different morphological types: VCC1632 (E), VCC 2095 (S0), VCC0596 (Sc). VCC1555(Sc), VCC1356(Sm), VCC1499(dE). 
The agreement is good  indicating once again the good quality of the photometry produced by our automatic pipeline. Nevertheless small differences appear
especially in late type galaxies in Fig. \ref{courteau}. We ascribe these effects to the fact that the profiles of \citet{mcdonald11} were extracted with an isophotal fitting procedure 
(i.e., ellipticity and PA of the ellipses are allowed to vary) contrary to ours. In fact the greatest deviations are displayed in VCC0596. VCC1555, VCC1356 that are three late type galaxies, respectively
an Sc, Sc and Sm. In these objects, isophotes twists in correspondence with non axisymmetric structures.

\subsection{Templates}
\label{guidotemplates}
This section is devoted to a more quantitative analysis of the color profiles and of their  correlations 
with stellar mass and morphology, focusing only on
Local supercluster galaxies,  which are resolved on scales of $\sim 100$ pc.
Moreover, in this way, we are not affected by the selection bias discussed in section \ref{secsample}.
Template profiles allow us to investigate the average properties of color profiles 
in our sample as a function of mass and morphology.
After normalizing each profile  to the Petrosian radius and
correcting for Galactic extinction, we create 
template profiles in different bins of stellar mass and morphology  with a radial step of $0.05$ R/R$_{{\it pet}}$.
Further on we correct the profiles for internal extinction. Despite the fact that we cannot obtain radial extinction profiles, \citet{dustyprof} 
show that on average the radial extinction profiles of spirals  are flat within the given errors, except for the very central region ($R\lsim 0.2~R_{25}$) where extinction can be 
significantly more severe with respect to the external parts. It
is impossible to correct for dust extinction  this region relying solely on the optical data and to implement a precise galaxy-to-galaxy dust correction for the whole sample is way beyond
the scope of the present study.
Therefore we correct the profiles applying the average correction evaluated for the total color
of the galaxy. In addition,  
we did not include galaxies with $incl>60^o$, in order to avoid contaminations from edge-on
spirals whose internal and external colors are dramatically reddened by the dust
extinction through the disk plane (see Fig. \ref{profils}d). 
We defined three equally populated mass bins:
\begin{description}
\item[-] a low mass bin containing galaxies with $M_*  \leq 10^{8.5}$ M$_{\odot}$,
\item[-] an intermediate mass bin, in which $10^{8.5} < M_* \leq 10^{10}$
  M$_{\odot}$,
\item[-] a high mass bin, in which $10^{10} <  M_*  \leq 10^{12}$ M$_{\odot}$.
\end{description}
In  the intermediate and high  mass bins we identified three morphological classes:
the first class contains elliptical galaxies, S0s and S0a;
the second bin includes late type galaxies from Sa to Sbc while the last bin  Sc and irregulars. 
The cut is based on the morphological classification taken from GOLDMine \citep{pg03,pg14b} which, in the Local Universe, is mainly based on
the morphological classification performed by \citet{rc3} and \citet{vcc} on photographic plates of exquisite quality.\\

Galaxies with masses lower than $10^{8.5}$ M$_{\odot}$
are either classified as dwarf ellipticals or dwarf irregulars.
The resulting template profiles are shown in Fig. \ref{tempmass}, separately
for each morphological bin. \\Summarizing, early type galaxies
do not show significant gradients, irrespective of their mass.  Late type
galaxies show instead flat profiles in the low mass bin while, at higher
masses, their profiles show evident color gradients, implying the presence of
a red component in their central region whose importance increases with
mass. Focusing on the highest mass bin, spirals are as red as ellipticals in
their central parts and as blue as dIrrs in their outskirts.  In other words,
given the well known color-sSFR relation, spirals are primarily quenched in a
central region whose importance is a function of mass while their outermost
regions remain star forming \citep{fox13}.

\section{Color-magnitude decomposition}
\label{colmagdecomp}
\begin{figure*} 
\begin{centering}
\includegraphics[scale=0.9]{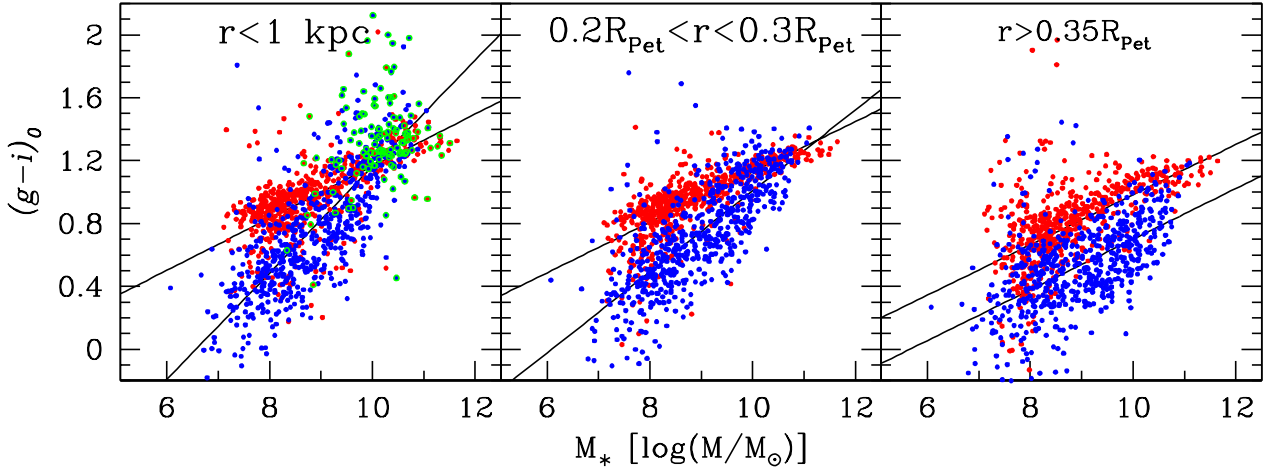}
\caption{Color-magnitude diagram of three different zones where red dots are 
ETGs and blue dots are LTGs: (left panel) inner zone, where open green dots are galaxies hosting an active galactic
nucleus of different kinds (LINERs, Seyfert, AGN) classified on the basis of nuclear 
emission lines; (central panel) intermediate zone; (right panel) outer, disk-dominated zone.}
\label{INOUT}
\end{centering}
\end{figure*}

Inspecting Fig. \ref{tempmass} with the aim of further investigating 
properties of color gradients, we identified three non-overlapping zones of interest of the color profiles:
\begin{description}
\item[-]the nuclear (or innermost) region
  going from the center of the galaxy up to approximately $1 \rm kpc$.
\item[-]an intermediate region defined by $0.2$R$_{Pet}\le$R$\le 0.3$R$_{Pet}$;
\item[-]an outer, disk-dominated, zone with R$\ge0.35$R$_{Pet}$.
\end{description}
Fig.~\ref{INOUT} shows the galaxy color as a function of the total stellar mass for each of the three 
zones, all corrected for Galactic extinction, 
with the disk-dominated zone also corrected for inclination. 
Overall, galaxies belonging to the red sequence are insensitive to this decomposition and show a consistent 
gradient in all three mass-color diagrams traced by each zone.
Instead late-type galaxies populate a blue cloud that show three different distributions 
for each of the three zone analyzed.
While the outer region of late type galaxies follows a relation almost parallel to the red sequence, the 
inner zones lie on much steeper color-mass relations. 
\label{templts}

\subsection{Nuclei}
\begin{figure} 
\begin{centering}
\includegraphics[scale=0.4]{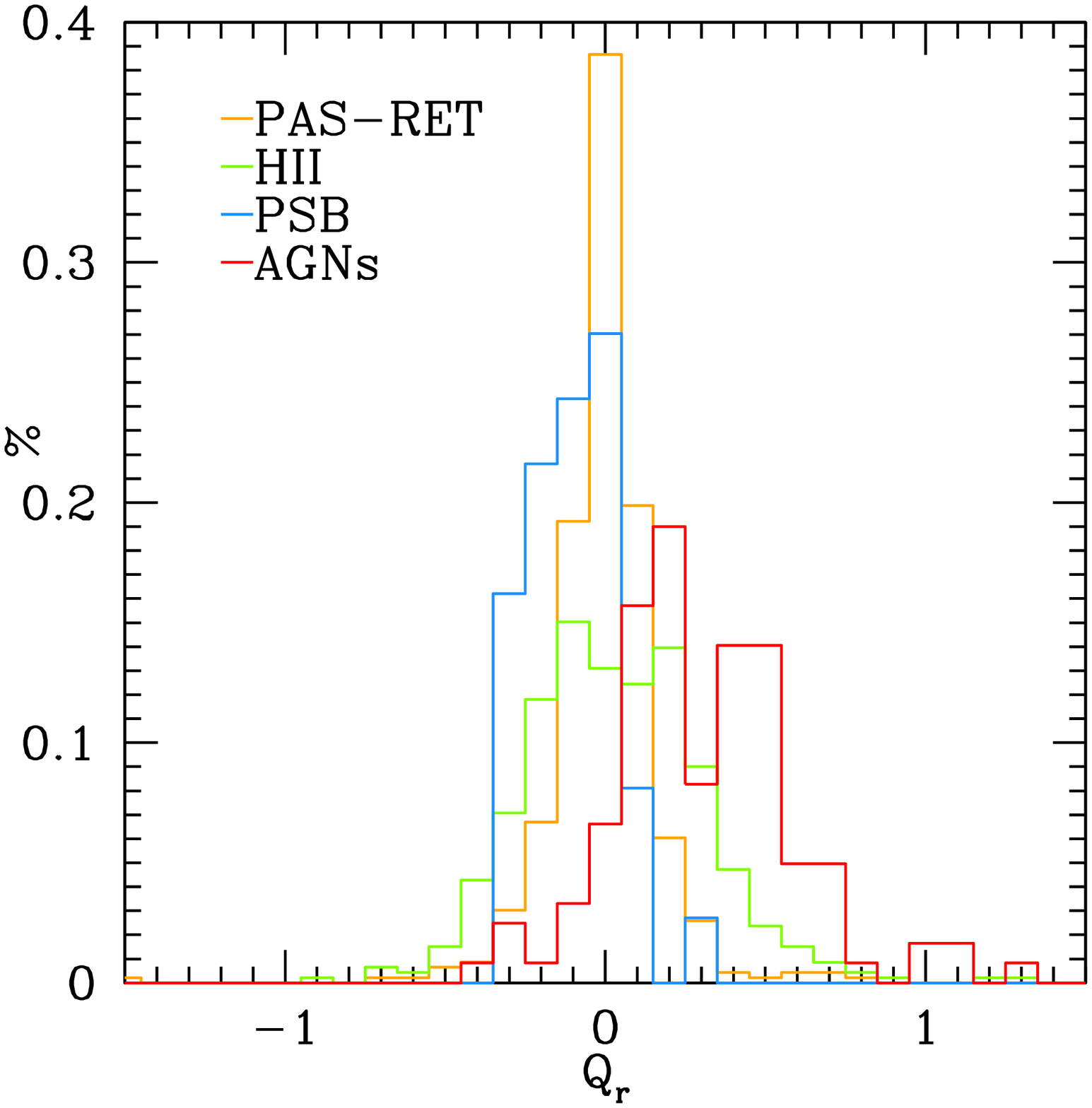}
\includegraphics[scale=0.4]{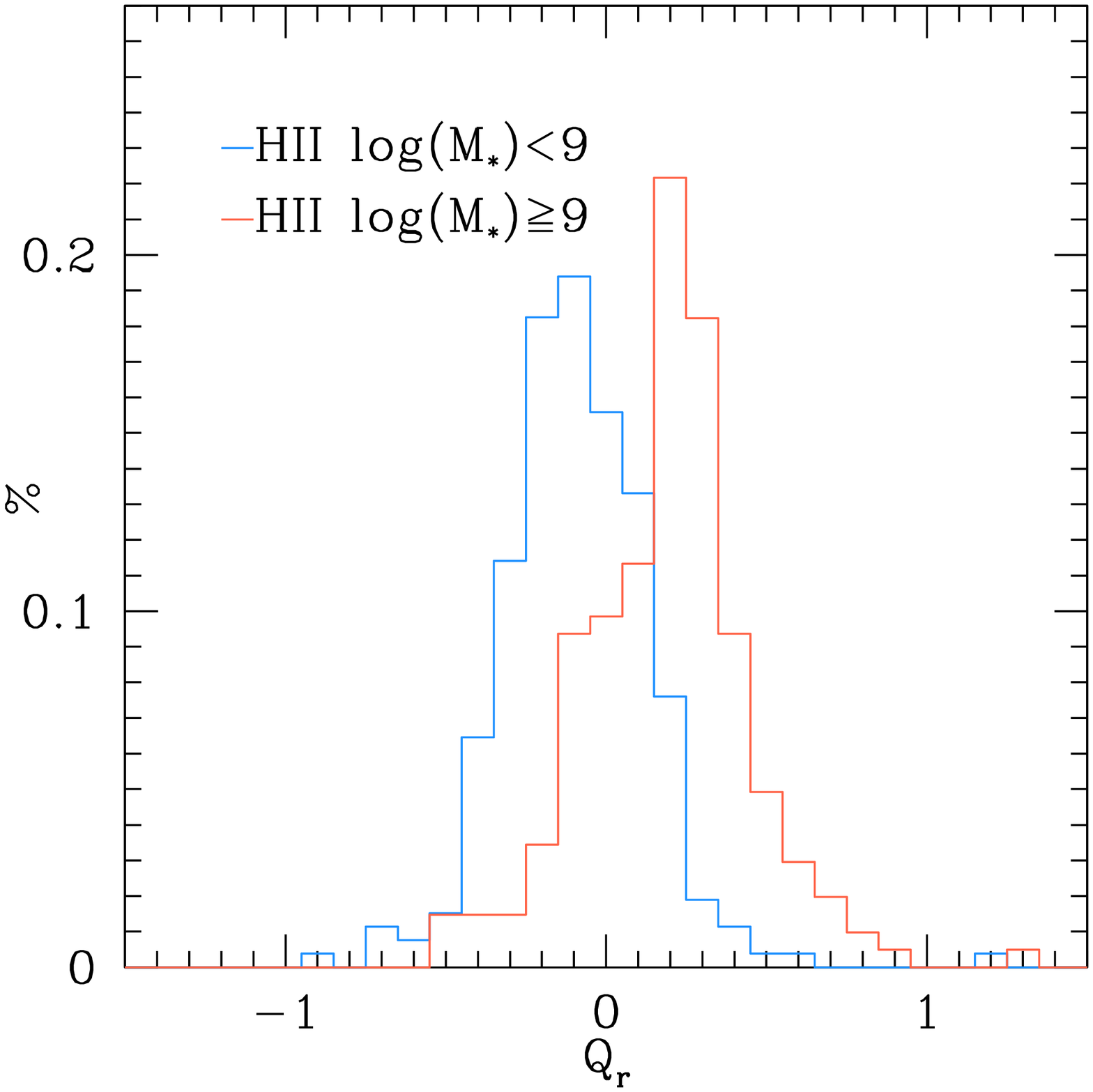}
\caption{(Top) Normalized distributions of the nuclear reddening Q$_r$ indices for  different classes of 
nuclear activity \citep[computed following][]{whan} color coded as follows: AGNs are represented by a red line;
the (PSBs) distribution is traced by the blu line; the green line stands for HII-like nuclei and the orange for PASSIVE and Retired galaxies.
(Bottom) Normalized distributions of HII-like nuclei above (red) and below (blue) 10$^9$M$_\odot$.}
\label{istnuc}
\end{centering}
\end{figure}

In the left panel of Fig.~\ref{INOUT} we plot the color mass diagram of the
inner \rm{kpc} of galaxies in our local supercluster sample.  Early type
galaxies form a well defined red sequence while late type galaxies lie on the
blue cloud. Comparing this diagram to the classical color-magnitude, the red
sequence slopes are consistent.  Nevertheless, considering nuclear colors,
late type galaxies  scatter around a much steeper relation that crosses
the red sequence at about $10^{9}$ M$_{\odot}$, displaying color indices even
greater than those for typical early types at M$_*>10^{9.5-10}$M$_{\odot}$. 
Furthermore, in the mass range $10^{9}$-$ 10^{12}$M$_{\odot}$, we highlight galaxies with ongoing
nuclear activity, such as AGNs, LINERS and Seyfert (green dots in the left
panel in Fig. \ref{INOUT}).  The adopted nuclear activity classification is
based on the ratio of nuclear H$\alpha$ and [NII] emission lines according to
the WHAN \citep{whan} diagram. Nuclear emission lines were taken from the SDSS
spectroscopic database Data Release 12 \citep{dr12}, complemented with spectra
available in NED and other measured by \citet{pg13c}.  Overall we find the nuclear classifications for $91\%$ of the galaxies.
 \\ As it can be seen in figure
\ref{INOUT}, these objects, highlighted in green, deviate from the relations
followed by the red sequence and the blue cloud and reach the most extreme
red values of color indices at any given mass. \\
To further investigate how the color inside the nuclear region is linked to its nuclear activity
we define an index of ``central/nuclear reddening`` :
\begin{equation}
Q_{red}=(g-i)_{nuc}-(g-i)_{0},
\end{equation}
where ($g$-$i$)$_{nuc}$ is the nuclear-scale color corrected for Galactic extinction and ($g$-$i$)$_{0}$ is the mean color of the galaxy. 
Q$_{red}$ estimates the deviation of the nuclear color from the average color of the galaxy.
Fig. \ref{istnuc} shows the  distribution of Q$_{red}$ for four different categories of nuclear activity: 
in orange, we plot the distribution of the RETIRED
\citep[galaxies that stopped their star formation and have their gas ionized by old stellar populations,][]{stasi15} and Passive nuclei galaxies;   
in red are displayed LINERs, AGNs and Seyfert; blue represents Post Staburst (PSB) galaxies and green is for 
HII region-like nuclei. \\
The identified categories distribute quite differently:
Passive and Retired galaxies show a quite tight distribution peaked at $Q_{\rm red}\sim 0$ ;
Active Galactic Nuclei are undoubtedly associated 
with colors that are redder than the average 
of the galaxy. 
The vast majority ($\sim91\%$) have values of $Q_{\rm red} > 0.0$ and only  $9 \%$ of AGNs display a bluer nucleus  with respect to the galaxy color.
Moreover, if we consider galaxies reaching the more extreme deviations, i.e. 
$Q_{\rm red} > 0.3$, they represent the 45\% of the AGNs population. This effect is possibly due to dust absorption 
associated with the gas fueling the nuclear region, although a contribution from the AGN itself cannot be excluded based only on the optical data.
Post Staburst (PSB) galaxies are, on the contrary, associated with nuclei that are bluer with respect to the 
galaxy color since in 80\% of cases PSBs display a $Q_{\rm red} < 0$;
HII-region like nuclei have instead the most widespread distribution of nuclear colors, 
with respect to the galaxy color. 
 The evidence of a bluer nucleus in PSBs and the fact that they are preferentially found in higher density environments is consistent with a picture
where these objects are transitioning  from the blue cloud to red sequence quenching their star formation in an outside-in fashion due to a ram pressure stripping event
\citep{pg10}.\\
Moreover, if we consider low and high mass galaxies separately (Fig.
\ref{istnuc}, {\it bottom panel}), the distribution is bimodal:  
 HII region-like nuclei in massive galaxies are always redder than the galaxy itself
while, at low masses, the distribution is dominated by galaxies showing a
nucleus that is bluer with respect to the average color of the
galaxy.\\
This is consistent with a picture where nuclei of small
galaxies are more likely to be caught in a bursting act and have little or no dust
absorption \citep{dustyprof} while, massive galaxies have on average a redder nucleus with respect to the galaxy color  possibly due to dust extinction.
Nevertheless, our optical data cannot exclude a contribution from an underlying older stellar population.
\subsection{Bulges, Bars and Disks}
Focusing on the second and third panel of Fig. \ref{INOUT}, we highlight the
intermediate and disk zone contribution to the color-magnitude diagram.    The
red sequence is once again consistent between the two plots.
Indeed, the slopes determined with a least squared fit of the red sequence in the two diagrams are consistently 
$0.17 \pm0.01 $ and $0.16 \pm 0.01$ mag~dex$^{-1}$. On the other hand, the blue cloud differs significantly: its slope varies from $0.34\pm 0.01$ mag~dex$^{-1}$ in the innermost 
region to $0.28 \pm 0.01$ and $0.18\pm 0.01$ mag~dex$^{-1}$ respectively in the intermediate and disk-dominated regions. 

Therefore, in the intermediate zone, the blue cloud displays a steep relation although  
not as steep as in the innermost region. Nevertheless, the blue cloud
completely overlaps the red sequence at M$_*>10^{10}$M$_{\odot}$.  This behavior
drastically changes  in the disk-dominated zone (right panel of
Fig. \ref{INOUT}).  Indeed, the outer-zone color magnitude diagram is  composed of
two well separated distributions: the blue cloud increases its color with
mass following a slope that is almost identical to the red sequence.\\
In the intermediate zone, above a threshold mass,  LTG galaxies have therefore suppressed SF and assume color values
typical of ETGs. On the contrary, they mostly appear as normal star forming objects in their outer disks. 

Moreover, the growth of the red component and the color contamination from the star forming disk (with bluer colors) in this zone 
can be approximately traced by the color difference between the intermediate and the outer zone  
shown in Fig.\ref{deltagi}, where this difference is plotted as a function of mass.
Despite a significant scatter, a Kolmogorov-Smirnov  test confirms
that the distributions of ETGs and LTGs are not drawn from the same parent sample (P$\lsim 10^{-3}$).
Red and blue lines connect the average difference between the intermediate and outer regions inside bins of 1 log(M$_*/$M$_{\odot}$) separately for 
ETGs (red) and LTGs (blue). 
This diagram shows that 
early type galaxies display an average mild gradient of approximately $\lsim 0.1$ mag from low to high mass galaxies.\\
Late type galaxies are instead characterized by a gradient between the two zones whose importance
increases with mass. This is consistent with a picture where galaxies develop a red and dead, quenched structure
in their central  regions, the relevance of which depends on mass. In their disks, instead, late type galaxies 
preserve their star formation almost unaffected. Nevertheless, even the outer  regions, which are occupied by SF 
structures and are therefore blue, still display an increasingly red color with increasing mass (see Fig \ref{INOUT}, right panel).
As a note of caution, we remind that the correction for internal extinction is not radially dependent and although it is a good approximation,
some small residual contribution of the dust absorption might still affect the profiles. Nevertheless, visual inspection of many images and profiles 
support the idea that the main cause of the red color observed is due to different stellar populations instead of dust extinction.  \\ 
 
\section{Discussion}
Our method successfully led to a set of final optical color-based parameters that trace
some properties of nuclei, bulges and disks among our Local sample.
In our analysis, we searched for systematic gradients in  $(g-i)$ color profiles in different galaxy types.
Here we compare our results with literature studies and discuss their implications for galaxy formation and evolution.\\

\noindent We investigated color gradients  with two different techniques: in section \ref{guidotemplates} we built color profiles templates for 
different morphological classes in three bins of increasing mass; in section 
\ref{colmagdecomp} we relate the galaxy total stellar mass versus the average color of three different, non-overlapping regions in which  
respectively nuclei, bulges (and/or bars) and disks are the dominant structure. As a general result, regardless of the binning used, 
colors show a mass dependency, with more massive galaxies harboring redder structures with respect to their lower mass counterparts (see Fig. \ref{INOUT}), confirming different literature results.  
\subsection{Nuclei}
Our investigation of nuclear regions confirms the tight correlation existing between the color of nuclei and the luminosity/mass of the host-galaxy
that has been consistently observed in detailed studies of massive Virgo cluster galaxies by the ACS Virgo cluster survey \citep{ACSvirgo} and of fainter galaxies
of the Virgo and Fornax clusters \citep{lotz04}.
We find that the correlation holds for both faint and bright galaxies although it shows a large scatter in the bright end of the distribution.
Consistently, \citet{ACSvirgo} have shown that the brightest early type galaxies show a considerable scatter in the nuclear color magnitude.
Nevertheless, a note of caution is required: the scale that defines our innermost region extends up to 1kpc, which is approximately ten times the scale length of the
nuclei studied in the ACS Virgo cluster survey and by \citet{lotz04} both relying on HST data. 
In particular, we note that our innermost region extends to the typical distance within which dusty structures such as the one in Fig. \ref{profil}(a) are found. 
In the ACS these represents $\sim 20\%$ of the ETGs and 
therefore at least for these early type galaxies, the high reddening is likely dominated by the presence of structures of dust
instead of their underlying stellar populations. 
Nevertheless we still find evidences of the results drawn by \citet{lotz04} and \citet{ACSvirgo}:
all nuclear regions of bright ETGs are redder than their harboring galaxy \citep{ACSvirgo} while, at 
the faint end of the red sequence, nuclear colors are more scattered and often bluer than the color of their host galaxy (see top panel of Fig. \ref{deltagi}).\\

LTGs show similar trends although strongly amplified compared to the ETGs. 
Intriguingly, among the high mass, scattered population of both ETGs and LTGs, it appears to be a correlation between the color in the innermost region 
and the AGN activity of the nucleus: nuclei of active galaxies are found on average to be redder  than their non-active counterparts (see Fig. \ref{INOUT}).
Nevertheless, given the large extension of our innermost region, we refrain from considering the AGN as main contributors to the color of the nuclei in these galaxies, 
as such effect could be related to the extinction caused by the dust dragged by the gas that occupies the region and fuels the AGN.
Our analysis is consistent with the stellar population study presented by \citet{ACSvirgo} who
find an old/intermediate stellar population component in the nuclear regions of all bright galaxies (ETGs and LTGs).
These still follow a color magnitude relation despite the scatter at high mass that suggests that,  at least for the most deviant population, the nuclear
chemical enrichment was governed by internal/local factors. 
The presence of disky dusty structures even in evolved systems such as Es and S0s suggests that the region is 
periodically refurbished with gas and dust. For example, secular disk instabilities or mergers can funnel the gas and the dust toward the very center of the galaxy.
Consequently the strong degeneracy between absorption and stellar population age on such scales prevents from a unique interpretation of the history of these structures 
relying solely on optical data \citep{driver07}.
Further on, as it is shown in the faint end of the two distributions in Fig \ref{deltagi} ({\it Top}), many LTGs have nuclei that are bluer compared to the galaxy
color \citep{taylor05} and, despite a considerable scatter, many ETGs also display blue nuclei consistently with \citet{lisk06}. Our spectroscopic analysis revealed that 
a fraction of the population of blue nuclei is possibly populated by PSB galaxies (see Fig \ref{istnuc})  
which have recently experienced a recent, sudden shut down of star formation and by nuclei that retain some residual SF activity  (HII-like) despite the poor HI content of the host
galaxy. These systems are undergoing an outside-in quenching of the star formation in the Virgo cluster hinting at 
physical processes such as harassment \citep{harassment} and ram pressure stripping \citep{ram} as possible causes of such evolutionary paths. 
\subsection{Bulges, Bars \& Disks}
\begin{figure} 
\begin{centering}
\includegraphics[scale=0.4]{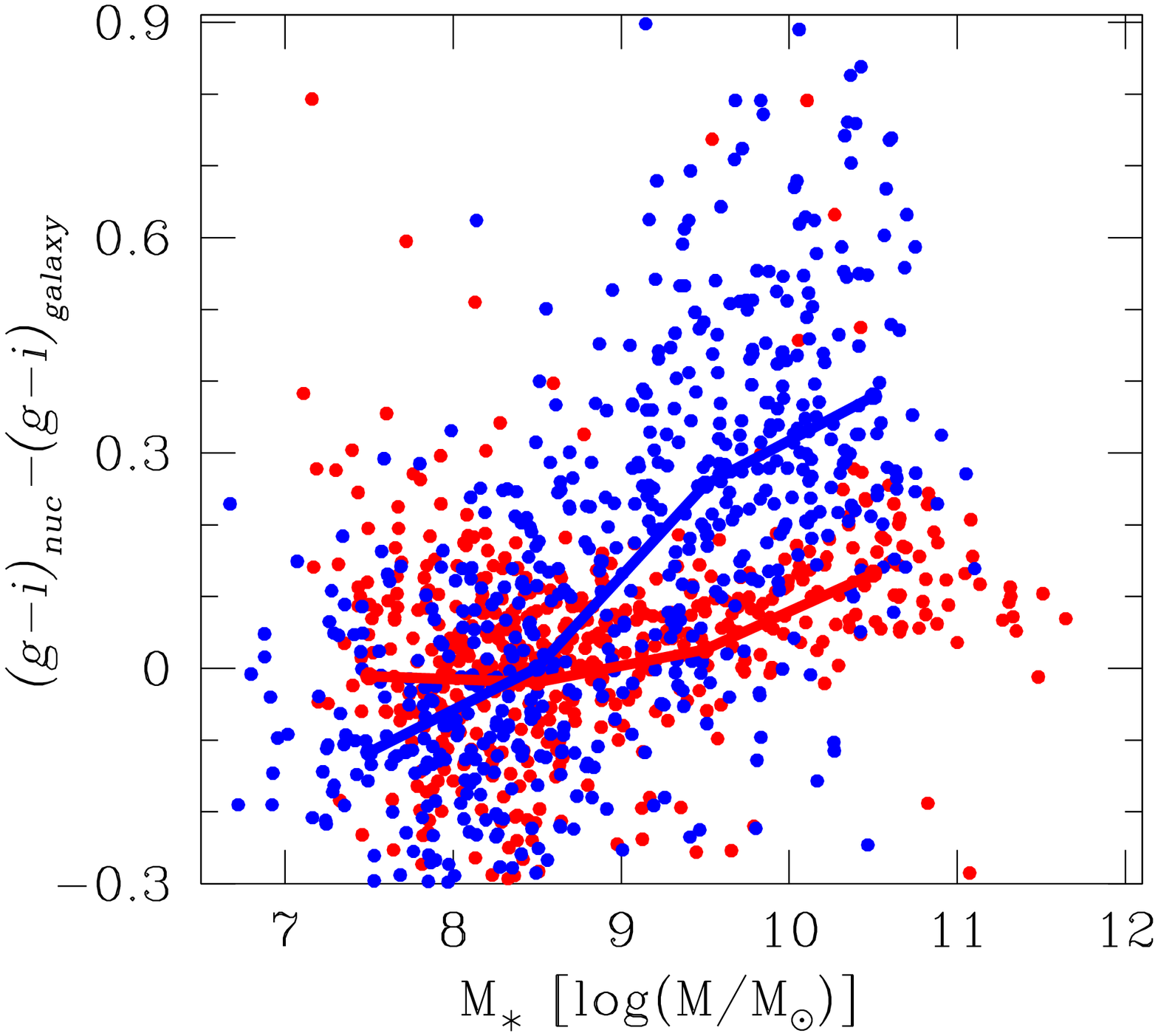}
\includegraphics[scale=0.4]{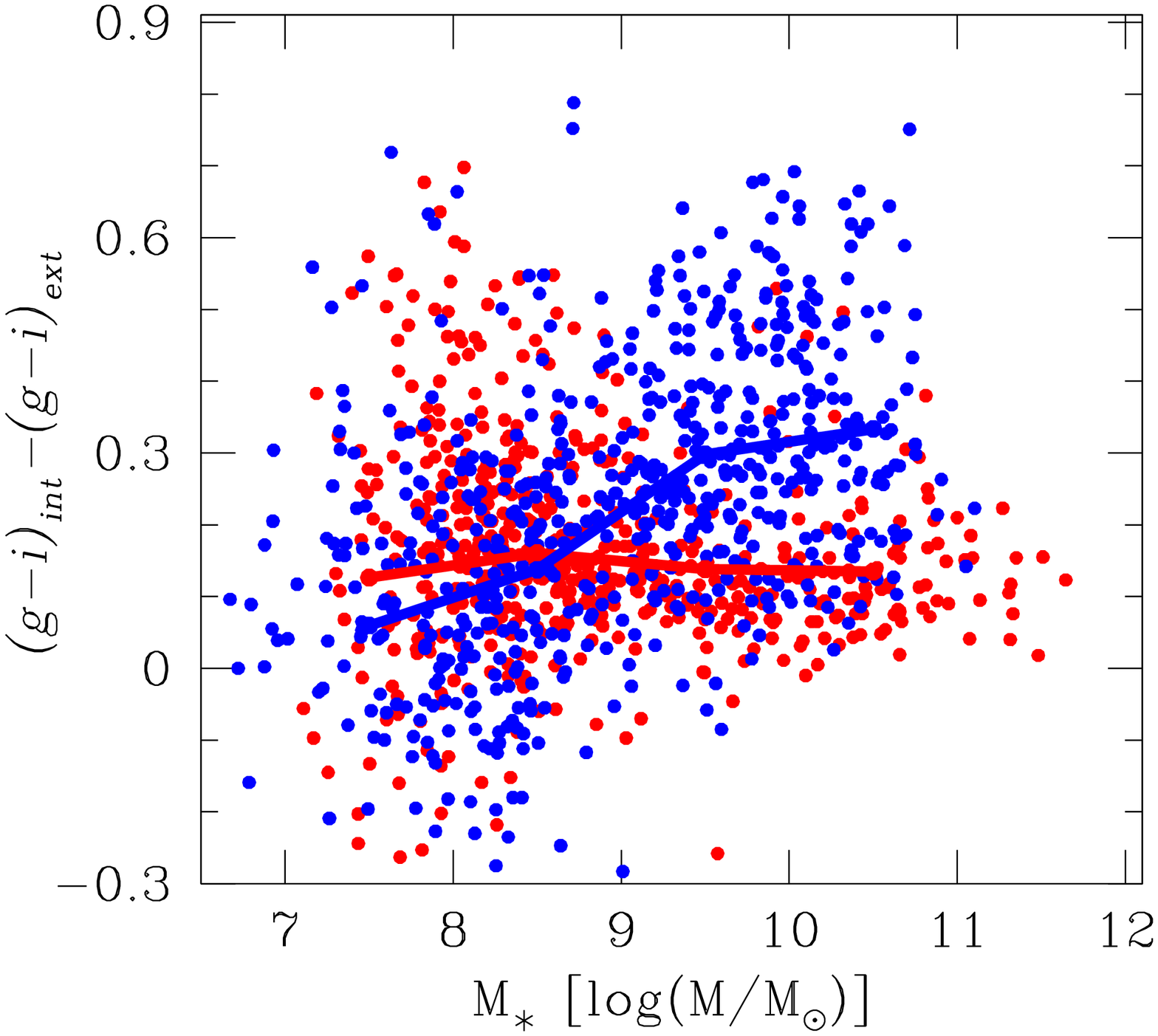}
\caption{ Top:Distribution of the difference between the color indexes 
of the nuclear zone and the color of the galaxy plotted against the total stellar mass.
Bottom:Distribution of the difference between the color indexes 
of the intermediate zone and those of the disk-dominated zone plotted against the total stellar mass.
In both panels, the blue and red small dots stand for respectively the blue cloud and the red sequence galaxies. The blue (red) line connects the average value in 4 bins of $10 \log(M/$M$_\odot)$ 
from $10^7$ to $10^{11}$ M$_\odot$ for LTGs (ETGs).}
\label{deltagi}
\end{centering}
\end{figure}

Despite the considerable scatter in both colors and color gradients \citep{balcells96, taylor05, roed_1}, the tight correlation between color and stellar mass of the host galaxies holds 
true in both the region identified as intermediate and outer in section \ref{colmagdecomp}. 
ETGs form a tight red sequence (see Fig. \ref{INOUT}) for both regions and show an average inside-out gradient of $\sim 0.1$ mag (Fig. \ref{deltagi}, {\it bottom}).
LTGs form instead two different distributions: the blue cloud of the intermediate (or bulge/bar) region becomes red (i.e., it reaches the red sequences) 
above $10^{10}$ M$_\odot$ while the outer, disk-dominated region never overlaps completely the red sequence. 
Still, more massive disks are redder than their lower mass counterparts but the difference between the colors of the outer and the intermediate region increases 
with respect to the total stellar mass. 
This can be possibly induced by the growth of a red and dead structure in the center of massive disks, i.e. that the central part of galaxies underwent a star formation quenching process 
that turned them red.
Our results on the average properties of color profiles broadly agree with literature data. For example \citet{macarth} have shown that the radial profile of the average ages of the
stellar populations decreases from inside out and that the steepness of the decrease is a function of morphological type.  
Color templates shown in Fig. \ref{tempmass} exhibit a radial behavior fully consistent with the average age profiles
shown by \citet{macarth}. 
We find there is also a good agreement with the (g-H) color profiles published in \citet{roed_1} for almost all the morphological types, although in their median profiles of early disks 
they find positive color gradients \citep[and consistently positive age population gradients in their stellar population analysis,][]{roed_2} that we do not see.
\citet{roed_2} do not find any direct link between galaxy morphologies and the observed stellar population gradients.
On the contrary, studies such as \citet{cheung2013} and \citet{pg15} have shown that the bar occupation fraction rises steeply above  $10^{9.5}$M$_\odot$ \citep[as also confirmed by the works done by ][]{Skibba12, Masters12} and that, 
above this mass,galaxies are progressively more quenched (red) in their centers, while their disks still sustain SF and hence are blue. These studies therefore highlight that 
the presence of structures such as bars can indeed produce the color gradients that we observe and likely also the stellar population gradients.
These authors  thus conclude that a secular bar drives the quenching of the star formation in the central kiloparsecs of galaxies. 
Moreover massive galaxies undergo bar instability earlier 
than their lower mass counterparts and thus have more time to grow redder than low mass systems. 
Moreover, \citet{abreubars} on 
a study of the Virgo bar fraction have shown that this rises up to more than $50\%$ above $10^{10}$M$_\odot$, adding a further link between color/stellar populations
radial gradients that we observe and structures such as bars (Laurikainen et al. 2010). 
We stress, however, that disk instabilities can also rejuvinate the central stellar population by, e.g., triggering central star formation 
in correspondence of the ILRs of spirals or bars (an example could be the one of VCC 508 in Fig. \ref{profil} (a)).\\
Both \citet{macarth} and \citet{roed_1}, also observe  a positive gradient in low mass galaxies. Among these galaxies we find only a mild gradient
in the template profile of dIrr in  Fig. \ref{tempmass}. From Fig. \ref{deltagi} we are taken to conclude that positive gradients can be found in low mass galaxies
especially if we consider the most internal regions \citep{lisk06, fox13} in contrast to the external parts but that the severe scattering occurring at low mass in our parameters 
prevents from extracting a robust general trend. This could be related to the fact that the low mass population may be composed of objects with different formation processes (van Zee et al
2004; Lisker et al. 2008). In our sample, positive gradients are most noticeable in the nuclear region (Fig. \ref{deltagi}, Top) while in the comparison between
the color of the intermediate region with  the color of the disk-dominated region their difference is consistent with zero.\\

\section{Summary and Conclusions}

In this paper we have presented a semi-automated, IDL-based, procedure designed to perform photometric extraction on SDSS multi-band 
images. The procedure was used to analyze a magnitude and volume limited sample of 5532 galaxies in the Local and Coma superclusters. 
Our procedure, unlike the SDSS pipeline, avoids the so-called shredding problem and successfully extracts total Petrosian 
magnitudes even for the largest objects of the sample, recovering 383 $g-$ (and $i-$) band magnitudes not included in the SDSS 
DR10 \citep{dr10}. After the galactic and internal extinction correction is applied, we recover a refined color-magnitude diagram 
of the Local and Coma superclusters. 

We attempted to dissect the color of galaxies in three distinct contributions: a nuclear region, an intermediate ($0.2$R$_{Pet}\le$R$\le 0.3$R$_{Pet}$), and an external region. 
To this  end, we analyzed the high quality 
color profiles that our routine has extracted for each galaxy, focusing on the objects of the Local Supercluster sample, 
with the best spatial resolution. Our analysis highlighted that:\\
Profiles are powerful tools for the detection of structures on nuclear scales (ILRs, dusty disks etc. ) even when they are difficult to be visually
spotted on SDSS images.\\
{\it i)} Nuclei of galaxies display the greatest deviations with respect to the galaxy color. This deviation correlates mildly with the nuclear activity of the galaxy.
Passive galaxies do not deviate from the color of the galaxy while AGNs generally display very red nuclear colors. Star forming nuclei follow instead a bimodal distribution: 
Galaxies that are more massive than $10^9$M$_\odot$ display redder nuclear color with respect to the galaxy while galaxies with $M<10^9$M$_\odot$ have bluer nuclei. 
Overall, a wide population of low mass galaxies display bluer nuclei with respect to their average color see Fig. \ref{deltagi} ({\it top}) although the high scatter
does not allow the extraction of a general trend .\\
{\it ii)} Profiles of spiral galaxies reveal on average an intermediate zone that is redder with respect of the outer disk and this component is more important at high masses.\\
{\it iii)} The intermediate zone color of LTGs  overlaps the red sequence already at $10^9$M$_\odot$ and is completely superposed to it at $10^{10}$M$_\odot$ (Fig. \ref{INOUT} and Fig. \ref{deltagi}).\\
{\it iv)} The disks of spiral galaxies follow a distribution that does not overlap (and is almost parallel to) with the red sequence but that still displays an increasing color 
index with increasing mass.\\

From {\it i)} we conclude that a wide population of low mass galaxies undergo an outside-in quenching of their star formation consistent with an evolution driven by the environment
on short timescales \citep{bos06,lisk06, fox13}. Nevertheless this conclusion does not hold for the whole population at low mass, as testified by the large scatter in their photometric parameters in both ETGs and
LTGs.
From {\it ii)} and {\it iii)} we conclude that massive spiral galaxies develop a red and dead component,  the importance of which 
increases with mass. Moreover, {\it i)} and {\it iii)} lead us to conclude that
this component must arise in a range mass between  $10^9$M$_\odot$ and $10^{10}$M$_\odot$ but that this component is not the only contribution to the slope of the global color-magnitude.
In fact, {\it iv)} demonstrates that even after subtracting the red and dead component another process \citep[such as strangulation,][]{peng15, fiaccons} 
is required to progressively quench the star formation in the outer regions of massive disks. \\

\begin{acknowledgements}
The authors would like to thank the referee, Thorsten Lisker, for 
his constructive criticism. This research has made use of the GOLDmine database (Gavazzi et
al. 2003, 2014b) and of the NASA/IPAC Extragalactic Database (NED) which is
operated by the Jet Propulsion Laboratory, California Institute of Technology,
under contract with the National Aeronautics and Space Administration.  We
wish to thank an unknown referee whose criticism helped improving the
manuscript.  Funding for the Sloan Digital Sky Survey (SDSS) and SDSS-II has
been provided by the Alfred P. Sloan Foundation, the Participating
Institutions, the National Science Foundation, the U.S. Department of Energy,
the National Aeronautics and Space Administration, the Japanese
Monbukagakusho, and the Max Planck Society, and the Higher Education Funding
Council for England.  The SDSS Web site is \emph{http://www.sdss.org/}.  The
SDSS is managed by the Astrophysical Research Consortium (ARC) for the
Participating Institutions.  The Participating Institutions are the American
Museum of Natural History, Astrophysical Institute Potsdam, University of
Basel, University of Cambridge, Case Western Reserve University, The
University of Chicago, Drexel University, Fermilab, the Institute for Advanced
Study, the Japan Participation Group, The Johns Hopkins University, the Joint
Institute for Nuclear Astrophysics, the Kavli Institute for Particle
Astrophysics and Cosmology, the Korean Scientist Group, the Chinese Academy of
Sciences (LAMOST), Los Alamos National Laboratory, the Max-Planck-Institute
for Astronomy (MPIA), the Max-Planck-Institute for Astrophysics (MPA), New
Mexico State University, Ohio State University, University of Pittsburgh,
University of Portsmouth, Princeton University, the United States Naval
Observatory, and the University of Washington.\\
M. Fossati acknowledges the support of the Deutsche Forschungsgemeinschaft via Project ID $3871/1-1$.
M. Fumagalli acknowledges support by the Science and Technology Facilities Council [grant number ST/L00075X/1]. 
\end{acknowledgements}

\begin{appendix}{}
\section{Source Extractor setup}
\label{setup}
Source Extractor (version $2.5.0$) remains one of the central tools of
the procedure. It has the crucial task to create the best possible mask for the field of view, crucial
to obtain a robust sky estimate,  and to evaluate the geometrical parameters of the target, fundamental to 
create the proper set of ellipses (section \ref{phot}). 
This is performed in the higher signal to noise white image (see section \ref{phot_reduction}).
The most challenging task is to guide Source Extractor to find efficiently small as well as 
large galaxies and properly de-blend bright contaminating sources minimizing the light loss from the target .\\

As our sample is dominated by galaxies belonging to the Coma supercluster, spanning a range of apparent dimensions 
from few tens of arc-seconds to few arc-minutes, 
we chose to favor a setup that works well with galaxies of this size and we implement an $ad-hoc$ procedure (described in 
appendix \ref{bigdeal}) for the few larger galaxies (all 
belonging to the Local supercluster), that extend for $\sim10$ arcmin and contain many resolved substructures.

For all other galaxies, we list here the best set of initial Source Extractor options that we adopted for the analysis.
In order to obtain a segmentation map for each image analyzed the $CHECKIMAGE\_TYPE$ parameter must be turned on $SEGMENTATION$.
Parameters that regulate the minimum number of pixel and the minimum flux limit for a detection are $DETECT\_MINAREA$ and 
$DETECT\_THRESH$ which we set respectively to 5 pixels and 1.5 sigmas ($THRESH\_TYPE =RELATIVE$ ). This parameters are the best to account for 
faint galaxies and galaxies affected by bright sources as also confirmed by \citet{Kim+14}. 
For all galaxies, the deblending parameters we adopted are  $DEBLEND\_MINCONT = 0.001$ and $DEBLEND\_NTHRESH  = 16$.
Further on, we modified the memory parameter $MEMORY\_PIXSTACK$ bringing it to $300000$.
Moreover, as we are only interested in the geometry of the target (PA, axis ratio and best pixel center coordinates),
we did not accurately tune parameters regulating the photometric extraction and the best sky estimate of Source Extractor.\\

\section{Dealing with large nearby galaxies}
\label{bigdeal}
Although our procedure is able to analyze galaxies that span a wide range of apparent dimensions, the Source Extractor
setup we adopted (Appendix \ref{setup}) fails for 3\% of the largest galaxies of the sample, all belonging to the Local supercluster.
In fact, galaxies belonging to the Local supercluster 
have many resolved substructures e.g., perturbed spiral arms, strongly obscuring dust lanes, 
bright HII regions etc.\\

These structures sometimes reach the deblending parameters (see appendix \ref{setup}) according to which Source Extractor separates overlapping objects. 
Therefore some substructures of the galaxy are detected by Source Extractor as entities not belonging to the target.

As a consequence, the area of the galaxy in the segmentation map is not composed by a single patch but, instead,
by as many patches as the different structures deblended by Source Extractor.\\
As described in section \ref{phot_reduction}, our procedure exploits the segmentation map in 
order to produce the mask for the image by removing
the central patch assuming that this is the only mask (of a galaxy, parameter $CLASS\_ STAR\sim 0$) (section \ref{phot_reduction}).
Therefore, in the case of the 3\% of problematic galaxies, our procedure would remove only a part of the mask covering the area 
occupied by the galaxy, leading to a wrong magnitude. 
To avoid the problem, the user must provide a list of the biggest galaxies, listing the following parameters:
\begin{description}
\item[-]{\bf \rm{RA}}  in degrees.
\item[-]{\bf \rm{Dec}}  in degrees.
\item[-]{\bf \rm{A}}, the major axis of the galaxy in {\rm arcmin}.
\item[-]{\bf \rm{B}}, the minor axis of the galaxy in {\rm arcmin}.
\item[-]{\bf \rm{PA}}, the position angle in degrees.
\end{description}
These parameters, taking advantage of the WCS system, are used to define the area occupied by the target
galaxy as the pixels inside the ellipse 
\begin{equation}
x=a \cos(\phi)\cos(P.A.)-b \sin(\phi)\sin(P.A.)  
\end{equation}
\begin{equation}
y=a \cos(\phi)\sin(P.A.)+b \sin(\phi)\cos(P.A.),
\end{equation}
where P.A. is the position angle, a and b respectively the semi-major and semi-minor axis and $\phi$ the azimuthal angle going from 0 to $2\pi$.
Within this area, the procedure removes all masks but the one that have been assigned by Source Extractor the $CLASS\_STAR$ parameter exceeding 0.8. 
\end{appendix}

\onecolumn
\newpage
\onecolumn
\scriptsize
\tiny
\begin{longtable}{cccccccccc}
\caption{Magnitudes extracted and corrected $(g-i)$ colors of the whole sample (one page sample)}\\
\hline\hline
 jName      & VCC      & CGCG  & NGC	& $R.A.~(J2000)$ & $DEC.~(J2000)$     & $g$   & $i$     &  $g-i_{mw0}$  &    $\log(M_*)$      \\
	    &	       &       &	& deg    	 & deg  	      & mag & mag 	&  mag          &  M$_\odot$    \\
 (1)	    & (2)      & (3)   & (4)	& (5)		 & (6)  	      & (7) & (8)  	&  (9)	        & (10)        \\	      
\hline 
\endfirsthead
\caption{continued.}\\
\hline\hline
 jName      & VCC      & CGCG  & NGC	& $R.A.~(J2000)$ & $DEC.~(J2000)$     & $g$   & $i$     &  $g-i_{mw0}$  &    $\log(M_*)$     \\
	    &	       &       &	& deg    	 & deg  	      & mag & mag 	&  mag          &  M$_\odot$    \\
 (1)	    & (2)      & (3)   & (4)	& (5)		 & (6)  	      & (7) & (8)  	&  (9)	        & (10)        \\	
\hline
\endhead
\hline
\endfoot   
 SDSSJ100016.4+244850 & 0  &   0       & 0     & 150.0685  &  24.81398   &  16.84  &   16.40  &    0.34  &    8.58  \\
 SDSSJ100215.0+255533 & 0  &   0       & 0     & 150.56271 &  25.9259	 &  17.95  &   17.47  &    0.42  &    8.23  \\
 SDSSJ100411.6+240924 & 0  &   0       & 0     & 151.04845 &  24.15678   &  17.28  &   16.46  &    0.65  &    8.75  \\
 SDSSJ100415.8+241416 & 0  &   0       & 0     & 151.06615 &  24.23782   &  15.40  &   14.41  &    0.78  &    9.80  \\
 SDSSJ100445.2+255832 & 0  &   0       & 0     & 151.18864 &  25.97567   &  17.27  &   16.35  &    0.78  &    9.04  \\
 SDSSJ100534.4+273051 & 0  &   0       & 0     & 151.39364 &  27.51437   &  15.93  &   15.37  &    0.40  &    9.03  \\
 SDSSJ100621.9+241424 & 0  &   0       & 0     & 151.59129 &  24.2401	 &  17.44  &   16.61  &    0.68  &    8.83  \\
 SDSSJ100659.1+274121 & 0  &   0       & 0     & 151.74643 &  27.68922   &  17.87  &   17.00  &    0.80  &    8.77  \\
\end{longtable}        
\normalsize            


\begin{thebibliography}{}
\bibitem[Aaronson et al.(1981)]{aar81} Aaronson, M., Persson, S.~E., \& Frogel, J.~A.\ 1981, \apj, 245, 18 
\bibitem[Abazajian et al.(2009)]{dr7} Abazajian, K.~N., et al.\ 2009, \apjs, 182, 543 
\bibitem[Adelman-McCarthy et al.(2006)]{dr4} Adelman-McCarthy, J.~K., Ag{\"u}eros, M.~A., Allam, S.~S., et al.\ 2006, \apjs, 162, 38
\bibitem[Ahn et al.(2014)]{dr10} Ahn, C.~P., Alexandroff, R., Allende Prieto, C., et al.\ 2014, \apjs, 211, 17 
\bibitem[Alam et al.(2015)]{dr12} Alam, S., Albareti, F.~D., Allende Prieto, C., et al.\ 2015, arXiv:1501.00963 
\bibitem[Baldwin et al.(1981)]{bpt} Baldwin, J.~A., Phillips, M.~M., \& Terlevich, R.\ 1981, \pasp, 93, 5 
\bibitem[Baldry et al.(2004)]{bal04} Baldry, I.~K., Glazebrook, K., Brinkmann, J., et al.\ 2004, \apj, 600, 681 
\bibitem[Bell et al.(2004)]{bel04} Bell, E.~F., Wolf, C., Meisenheimer, K., et al.\ 2004, \apj, 608, 752 
\bibitem[Berriman et al.(2004)]{montage} Berriman, G.~B., Good,J.~C., Laity, A.~C., et al.\ 2004, Astronomical Data Analysis Software and Systems (ADASS) XIII, 314, 593 
\bibitem[Bertin \& Arnouts(1996)]{sex96} Bertin, E., \& Arnouts, S.\ 1996, \aaps, 117, 393 
\bibitem[Binggeli et al.(1985)]{vcc} Binggeli, B., Sandage, A., \& Tammann, G.~A.\ 1985, \aj, 90, 1681 
\bibitem[Blanton et al.(2001)]{bla01a} Blanton, M.~R., Dalcanton, J., Eisenstein, D., et al.\ 2001, \aj, 121, 2358 
\bibitem[Blanton et al.(2003)]{blanton03} Blanton, M.~R., Lin, H., Lupton, R.~H., et al.\ 2003, \aj, 125, 2276 
\bibitem[Blanton et al.(2005)]{blanton05} Blanton, M.~R., Schlegel, D.~J., Strauss, M.~A., et al.\ 2005, \aj, 129, 2562 
\bibitem[Blanton et al.(2011)]{blanton11} Blanton, M.~R., Kazin, E., Muna, D., Weaver, B.~A., \& Price-Whelan, A.\ 2011, \aj, 142, 31 
\bibitem[{{Boselli} \& {Gavazzi}(2006)}]{bos06} {Boselli}, A. \& {Gavazzi}, G. 2006, \pasp, 118, 517
\bibitem[Boselli et al.(2011)]{guvics} Boselli, A., Boissier, S., Heinis, S., et al.\ 2011, \aap, 528, A107 
\bibitem[Boselli et al.(2016)]{1690} Boselli, A., Cuillandre, J.~C., Fossati, M., et al.\ 2016, \aap, 587, A68
\bibitem[Chester \& Roberts(1964)]{ches64} Chester, C., \& Roberts, M.~S.\ 1964, \aj, 69, 635
\bibitem[Cheung et al.(2013)]{cheung2013} Cheung, E., Athanassoula, E., Masters, K.~L., et al.\ 2013, \apj, 779, 162  
\bibitem[Cid Fernandes et al.(2011)]{whan} Cid Fernandes, R., Stasi{\'n}ska, G., Mateus, A., \& Vale Asari, N.\ 2011, \mnras, 413, 1687 
\bibitem[Comer{\'o}n et al.(2010)]{comer10} Comer{\'o}n, S., Knapen, J.~H., Beckman, J.~E., et al.\ 2010, \mnras, 402, 2462 
\bibitem[C{\^o}t{\'e} et al.(2006)]{ACSvirgo} C{\^o}t{\'e}, P., Piatek, S., Ferrarese, L., et al.\ 2006, \apjs, 165, 57
\bibitem[Cotini et al.(2013)]{cotini13} Cotini, S., Ripamonti, E., Caccianiga, A., et al.\ 2013, \mnras, 431, 2661 
\bibitem[de Vaucouleurs et al.(1991)]{rc3} de Vaucouleurs, 
G., de Vaucouleurs, A., Corwin, H.~G., Jr., et al.\ 1991, Third Reference 
Catalogue of Bright Galaxies.~Volume I: Explanations and references.~ 
Volume II: Data for galaxies between 0$^{h}$ and 12$^{h}$.~ Volume III: 
Data for galaxies between 12$^{h}$ and 24$^{h}$., by de Vaucouleurs, G.; de 
Vaucouleurs, A.; Corwin, H.~G., Jr.; Buta, R.~J.; Paturel, G.; Fouqu{\'e}, 
P..~Springer, New York, NY (USA), 1991, 2091 p., ISBN 0-387-97552-7, Price 
US\$ 198.00.~ISBN 3-540-97552-7, Price DM 448.00.~ISBN 0-387-97549-7 
(Vol.~I), ISBN 0-387-97550-0 (Vol.~II), ISBN 0-387-97551-9 (Vol.~III).,  
\bibitem[Driver et al.(2007)]{driver07} Driver, S.~P., Popescu, C.~C., Tuffs, R.~J., et al.\ 2007, \mnras, 379, 1022 
\bibitem[Faber(1973)]{faber73} Faber, S.~M.\ 1973, \apj, 179, 731 
\bibitem[Fiacconi et al.(2015)]{fiaccons} Fiacconi, D., Feldmann, R., \& Mayer, L.\ 2015, \mnras, 446, 1957 
\bibitem[Font et al.(2014)]{font14} Font, J., Beckman, J.~E., Querejeta, M., et al.\ 2014, \apjs, 210, 2 
\bibitem[Fossati et al.(2013)]{fox13} Fossati, M., Gavazzi, G., Savorgnan, G., et al.\ 2013, \aap, 553, AA91 
\bibitem[Fumagalli et al.(2014)]{fuma14} Fumagalli, M., O'Meara, J.~M., Prochaska, J.~X., Kanekar, N., \& Wolfe, A.~M.\ 2014, \mnras, 444, 1282 
\bibitem[Gavazzi (1993)]{pg93} Gavazzi, G.\ 1993, \apj, 419, 469 
\bibitem[Gavazzi et al.(1996)]{pg96} Gavazzi, G., Pierini, D., \& Boselli, A.\ 1996, \aap, 312, 397 
\bibitem[Gavazzi et al.(1999)]{pg99} Gavazzi, G., Boselli, A., Scodeggio, M., Pierini, D., \& Belsole, E.\ 1999, \mnras, 304, 595 
\bibitem[Gavazzi et al.(2003)]{pg03} Gavazzi, G., Boselli, A., Donati, A., Franzetti, P., \& Scodeggio, M.\ 2003, \aap, 400, 451 
\bibitem[Gavazzi et al.(2008)]{pg08} Gavazzi, G., Giovanelli, R., Haynes, M.~P., et al.\ 2008, \aap, 482, 43 
\bibitem[Gavazzi(2009)]{pg09} Gavazzi, G.\ 2009, Revista Mexicana de Astronomia y Astrofisica Conference Series, 37, 72 
\bibitem[Gavazzi et al.(2010)]{pg10} Gavazzi, G., Fumagalli, M., Cucciati, O., \& Boselli, A.\ 2010, \aap, 517, A73 
\bibitem[Gavazzi et al.(2012)]{pg12} Gavazzi, G., Fumagalli, M., Galardo, V., Grossetti, F.,  Boselli, A., Giovanelli, R., Haynes, M.~P. \& Fabello, S., Paper I, \ 2012, \aap, 545, A16 
\bibitem[Gavazzi et al.(2013a)]{pg13a} Gavazzi, G., Fumagalli, M., Galardo, V., Grossetti, F.,  Boselli, A., Giovanelli, R.\& Haynes, M.~P., Paper II, 2013a,  \aap, 553, A89
\bibitem[Gavazzi et al.(2013b)]{pg13b} Gavazzi, G., Savorgnan, G., Fossati, M., et al. Paper III, \ 2013b, \aap, 553, AA90 
\bibitem[Gavazzi et al.(2013c)]{pg13c} Gavazzi, G., Consolandi, G., Dotti, M., et al.\ 2013c, \aap, 558, AA68 
\bibitem[Gavazzi et al.(2014b)]{pg14b} Gavazzi, G., Franzetti, P., \& Boselli, A.\ 2014b, arXiv:1401.8123 
\bibitem[Gavazzi et al.(2015)]{pg15} Gavazzi, G., Consolandi, G., Dotti, M., et al.\ 2015, \aap, 580, A116
\bibitem[Gawiser et al.(2006)]{gawiser} Gawiser, E., van Dokkum, P.~G., Herrera, D., et al.\ 2006, \apjs, 162, 1 
\bibitem[Graham \& Driver(2005)]{kron} Graham, A.~W., \& Driver, S.~P.\ 2005, \pasa, 22, 118
\bibitem[Griersmith(1980)]{grier80} Griersmith, D.\ 1980, \aj, 85, 1135 
\bibitem[Gunn \& Gott(1972)]{ram} Gunn, J.~E., \& Gott, J.~R., III 1972, \apj, 176, 1
\bibitem[Haynes et al.(2011)]{alfa40} Haynes, M.~P., Giovanelli, R., Martin, A.~M., et al.\ 2011, \aj, 142, 170
\bibitem[Hogg et al.(2004)]{hogg04} Hogg, D.~W., Blanton, M.~R., Brinchmann, J., et al.\ 2004, \apjl, 601, L29 
\bibitem[Holwerda et al.(2005)]{dustyprof} Holwerda, B.~W., Gonzalez, R.~A., Allen, R.~J., \& van der Kruit, P.~C.\ 2005, \aj, 129, 1396
\bibitem[Katz et al. (2011)]{k11}  Katz, D. S.,  Berriman, G. B. Mann, R. G. in Reshaping Research and Development Using Web 2.0-based Technologies. Editor: Mark Baker, 
Nova Science Publishers, Inc.(2011)
\bibitem[Kauffmann et al.(2004)]{kauff04} Kauffmann, G., White, S.~D.~M., Heckman, T.~M., et al.\ 2004, \mnras, 353, 713 
\bibitem[Kim et al.(2014)]{Kim+14} Kim, S., Rey, S.-C., Jerjen, H., et al.\ 2014, \apjs, 215, 22 
\bibitem[Kormendy \& Kennicutt(2004)]{K&K04} Kormendy, J., \& Kennicutt, R.~C., Jr.\ 2004, \araa, 42, 603 
\bibitem[Landsman(1993)]{1993ASPC...52..246L} Landsman, W.~B.\ 1993, Astronomical Data Analysis Software and Systems II, 52, 246 
\bibitem[Laurikainen et al.(2010)]{lau10} Laurikainen, E., Salo, H., Buta, R., Knapen, J.~H., \& Comer{\'o}n, S.\ 2010, \mnras, 405, 1089 
\bibitem[Lisker et al.(2006)]{lisk06} Lisker, T., Glatt, K., Westera, P., \& Grebel, E.~K.\ 2006, \aj, 132, 2432
\bibitem[Lisker et al.(2008)]{lisk08} Lisker, T., Grebel, E.~K., \& Binggeli, B.\ 2008, \aj, 135, 380
\bibitem[Lotz et al.(2004)]{lotz04} Lotz, J.~M., Miller, B.~W., \& Ferguson, H.~C.\ 2004, \apj, 613, 262 
\bibitem[Lupton et al.(2001)]{pipeline} Lupton, R., Gunn, J.~E., Ivezi{\'c}, Z., Knapp, G.~R., \& Kent, S.\ 2001, Astronomical Data Analysis Software and Systems X, 238, 269 
\bibitem[MacArthur et al.(2004)]{macarth} MacArthur, L.~A., Courteau, S., Bell, E., \& Holtzman, J.~A.\ 2004, \apjs, 152, 175 
\bibitem[Masters et al.(2012)]{Masters12} Masters, K.~L., Nichol, R.~C., Haynes, M.~P., et al.\ 2012, \mnras, 424, 2180 
\bibitem[McDonald et al.(2011)]{mcdonald11} McDonald, M.,Courteau, S., Tully, R.~B., \& Roediger, J.\ 2011, \mnras, 414, 2055 
\bibitem[Mei et al.(2007)]{mei07} Mei, S., Blakeslee, J.~P., C{\^o}t{\'e}, P., et al.\ 2007, \apj, 655, 144 
\bibitem[Melvin et al. (2014)]{} Melvin et al. \ 2014, arXiv:1401.3334 
\bibitem[M{\'e}ndez-Abreu et al.(2012)]{abreubars} M{\'e}ndez-Abreu, J., S{\'a}nchez-Janssen, R., Aguerri, J.~A.~L., Corsini, E.~M., \& Zarattini, S.\ 2012, \apjl, 761, L6 
\bibitem[Micheva et al.(2013)]{minc13} Micheva, G., {\"O}stlin, G., Bergvall, N., et al.\ 2013, \mnras, 431, 102
\bibitem[Moore et al.(1996)]{harassment} Moore, B., Katz, N., Lake, G., Dressler, A., \& Oemler, A.\ 1996, \nat, 379, 613 
\bibitem[Nair \& Abraham(2010)]{2010ApJ...714L.260N} Nair, P.~B., \& Abraham, R.~G.\ 2010, \apjl, 714, L260 
\bibitem[Peletier \& Balcells(1996)]{balcells96} Peletier, R.~F., \& Balcells, M.\ 1996, \aj, 111, 2238 
\bibitem[Peng et al.(2015)]{peng15} Peng, Y., Maiolino, R., \& Cochrane, R.\ 2015, \nat, 521, 192 
\bibitem[Roediger et al.(2011a)]{roed_1} Roediger, J.~C., Courteau, S., McDonald, M., \& MacArthur, L.~A.\ 2011, \mnras, 416, 1983
\bibitem[Roediger et al.(2011b)]{roed_2} Roediger, J.~C., Courteau, S., MacArthur, L.~A., \& McDonald, M.\ 2011, \mnras, 416, 1996 
\bibitem[Solanes et al.(1996)]{incl} Solanes, J.~M., Giovanelli, R., \& Haynes, M.~P.\ 1996, \apj, 461, 609 
\bibitem[Schlegel et al.(1998)]{dust98} Schlegel, D.~J., Finkbeiner, D.~P., \& Davis, M.\ 1998, \apj, 500, 525 
\bibitem[Schlafly \& Finkbeiner(2011)]{dust11} Schlafly, E.~F., \& Finkbeiner, D.~P.\ 2011, \apj, 737, 103 
\bibitem[Skibba et al.(2012)]{Skibba12} Skibba, R.~A., Masters, K.~L., Nichol, R.~C., et al.\ 2012, \mnras, 423, 1485
\bibitem[Stasi{\'n}ska et al.(2015)]{stasi15} Stasi{\'n}ska, G., Costa-Duarte, M.~V., Vale Asari, N., Cid Fernandes, R., \& Sodr{\'e}, L.\ 2015, \mnras, 449, 559 
\bibitem[Strateva et al.(2001)]{strat01} Strateva, I., Ivezi{\'c}, {\v Z}., Knapp, G.~R., et al.\ 2001, \aj, 122, 1861 
\bibitem[Strauss et al.(2002)]{pipeline2} Strauss, M.~A., Weinberg, D.~H., Lupton, R.~H., et al.\ 2002, \aj, 124, 1810
\bibitem[Tamura \& Ohta(2003)]{tamura03} Tamura, N., \& Ohta, K.\ 2003, \aj, 126, 596 
\bibitem[Taylor et al.(2005)]{taylor05} Taylor, V.~A., Jansen, R.~A., Windhorst, R.~A., Odewahn, S.~C., \& Hibbard, J.~E.\ 2005, \apj, 630, 784 
\bibitem[van Zee et al.(2004)]{vanzee04} van Zee, L., Barton, E.~J., \& Skillman, E.~D.\ 2004, \aj, 128, 2797
\bibitem[Visvanathan \& Sandage(1977)]{visva77} Visvanathan, N., \& Sandage, A.\ 1977, \apj, 216, 214 
\bibitem[Visvanathan \& Griersmith(1977)]{visva77b} Visvanathan, N., \& Griersmith, D.\ 1977, \aap, 59, 317 
\bibitem[Wu et al.(2005)]{wu05} Wu, H., Shao, Z., Mo, H.~J., Xia, X., \& Deng, Z.\ 2005, \apj, 622, 244
\bibitem[Xue et al.(2010)]{hzcmag} Xue, Y.~Q., Brandt, W.~N., Luo, B., et al.\ 2010, \apj, 720, 368\bibitem[Yasuda et al.(2001)]{yasu01} Yasuda, N., Fukugita, M., Narayanan, V.~K., et al.\ 2001, \aj, 122, 1104 
\bibitem[York et al.(2000)]{2000AJ....120.1579Y} York, D.~G., Adelman, J., Anderson, J.~E., Jr., et al.\ 2000, \aj, 120, 1579 
\bibitem[Zibetti et al.(2009)]{zibi09} Zibetti, S., Charlot, S., \& Rix, H.-W.\ 2009, \mnras, 400, 1181 
\bibitem[Zwicky et al.(1968)]{cgcg} Zwicky, F., Herzog, E., \& Wild, P.\ 1968, Pasadena: California Institute of Technology (CIT), 1961-1968,  
\end{thebibliography}
\end{document}